\documentclass[final,3p,times,twocolumn]{elsarticle}

\usepackage{amssymb}
\biboptions{sort&compress}

\journal{Physics Letters B}

\usepackage{amsmath,amssymb,amsfonts,graphicx,natbib}
\usepackage{bm}
\usepackage{slashed}
\usepackage{subfigure}
\usepackage[svgnames]{xcolor}
\usepackage[section]{placeins}

\def\SF{{\cal P}_\mathrm{SF}}
\def\vfr{$\varphi_R$}
\def\fr{\varphi_R}
\def\NL{N_L}
\def\qp{q^{\uparrow}}
\newcommand{\vect}[1]{\boldsymbol{#1}}
\def\mathbi#1{\textbf{\em #1}}

\newcommand{\al}[1]{\begin{align} #1 \end{align}}
\newcommand{\non}{\nonumber}
\newcommand{\vf}{\varphi}

\newcommand{\Gs}{\mathrm{GeV}^2}
\newcommand{\ImS}{0.47\columnwidth}
\newcommand{\ImM}{0.8\columnwidth}
\newcommand{\ImL}{0.8\columnwidth}
\begin{document}

\begin{frontmatter}

\title{Studies of Azimuthal Modulations in Two Hadron Fragmentation of a Transversely Polarised Quark}


\author[ADR_CSSM]{Hrayr~H.~Matevosyan}
\author[ADR_EVN,ADR_INFN]{Aram~Kotzinian}
\author[ADR_CSSM]{Anthony~W.~Thomas}

\address[ADR_CSSM]
{
CSSM and ARC Centre of Excellence for Particle Physics at the Tera-scale,\\ 
School of Chemistry and Physics, \\
University of Adelaide, Adelaide SA 5005, Australia
\\ http://www.physics.adelaide.edu.au/cssm
}
\address[ADR_EVN]
{
Yerevan Physics Institute,
2 Alikhanyan Brothers St.,
375036 Yerevan, Armenia
}
\address[ADR_INFN]
{
INFN, Sezione di Torino, 10125 Torino, Italy
}

\begin{abstract}

 We study the azimuthal modulations of dihadron fragmentation functions (DiFFs) of a transversely polarised quark using an NJL-jet based model that incorporates the  Collins effect for single hadron emission. The DiFFs are extracted as Monte Carlo (MC) averages of corresponding multiplicities using their probabilistic interpretation. To simplify the model and highlight the possible mechanisms that create this modulation, we choose the elementary Collins function to be proportional to the elementary unpolarised fragmentation and a constant probability ($\SF$) for the quark to flip its spin after a single hadron emission. Moreover, as a leading order calculation, only one of the produced hadrons in the decay chain of the quark is produced with elementary Collins modulation. We calculate  the dependence of the polarised DiFFs on various angles such as the azimuthal angle of the single hadron and the angle of the two hadron production plane \vfr\ for several values of $\SF$. We observe that the polarised DiFFs for oppositely charged pion pairs exhibit a $\sin($\vfr$)$ modulation. This effect is induced purely via the elementary Collins effect and persists even when the quark completely depolarises after a single hadron emission ($\SF=0.5$). Moreover, similar sine modulations are present in the distribution of pion pairs with respect to the azimuthal angle of their total transverse momentum, $\vf_T$.
\end{abstract}




\end{frontmatter}

\section{Introduction}

 The study of parton number, momentum and spin distributions inside the nucleon remains one of the most important topics in hadronic physics. Various unpolarised and polarised parton distribution functions (PDFs) encode these quantities and are extensively studied in both theory and experiment. There are only three collinear PDFs describing the nucleon at leading twist approximation: unpolarised, helicity and transversity PDFs. While the first two have been extensively studied and parametrizations fitted to data from various Deep Inelastic Scattering (DIS) experiments: both fully- and semi-inclusive, the transversity PDF is relatively poorly constrained because of its chiral-odd nature. Over the past several years a great deal of effort has been concentrated in studying the transversity PDF using two approaches. The first involves semi-inclusive DIS (SIDIS) measurements on a transversely polarised nucleon target, with a single hadron recorded in the final state. In the corresponding single spin asymmetry (SSA) the transversity is convoluted with the so-called Collins fragmentation function (FF). The second approach relies on SIDIS measurements with a transversely polarised nucleon target, where two hadrons are recorded in the final state. Here the SSA has a term, where transversity is multiplied by a so-called interference dihadron fragmentation function (IFF), that can be accessed by measuring the asymmetries in two back to back hadron pair production in $e^+e^-$ annihilation~\cite{Artru:1995zu,Boer:2003ya}.  These terms involving the transversity PDFs in both one- and two- hadron SIDIS are extracted experimentally using their sine modulations with the so-called Collins and \vfr\ angles respectively~\cite{Collins:1993kq,Jaffe:1997hf, Bianconi:1999cd, Bianconi:1999uc, Radici:2001na,Bacchetta:2003vn} . Transversity is then extracted by either modelling or parametrizing all the remaining functions entering the measured SSAs: unpolarised PDF, Collins and unpolarised FF for single hadron SIDIS, and the IFF and unpolarised dihadron fragmentation function (DiFF) for the two hadron case.  Moreover, it is believed that the two SSAs are generated by different mechanisms, namely the Collins effect in one-hadron SIDIS and the interference of the hadron pair production amplitudes in two hadron SIDIS. Recently the COMPASS collaboration presented the results of their analysis demonstrating a similarity between SSAs extracted with these two methods~\cite{Adolph:2014fjw,COMPASS:2013combo}. Namely, they found that the SSA for pairs of oppositely charged hadrons appears to be very close to the Collins asymmetry for positively charged hadron production, which in turn is very close to that for negatively charged hadrons taken with opposite sign.  Also,  both~\vfr~and $\vf_T$~modulations have been suggested to occur because of the Collins effect in Refs.~\cite{Artru:1995zu,Artru:2002pua}. Further, both the unpolarized DiFF and the IFF  at large invariant mass were recently calculated using perturbative quantum chromodynamics in Ref.~\cite{Zhou:2011ba}. Here it was shown that the IFF in the large invariant mass regime is intimately connected to the Collins fragmentation function at large transverse momentum.

  The dihadron approach has recently attracted a lot of attention, with the first extraction of transversity performed in Ref.~\cite{Bacchetta:2011ip,Bacchetta:2012ty} using the SIDIS two hadron SSA measured by HERMES~\cite{Airapetian:2008sk} and COMPASS~\cite{Adolph:2012nw}, along with $e^+e^-$  measurements by the BELLE collaboration~\cite{Vossen:2011fk}. IFFs and unpolarised DiFFs were extracted from fits to BELLE data either using spectator model calculations~\cite{Bacchetta:2006un,Bacchetta:2008wb} or parametric forms~\cite{Courtoy:2012ry}, along with input from Monte Carlo (MC) unpolarised event generator PYTHIA. 
    
  Here we study the dihadron fragmentation function to oppositely charged pions for a transversely polarised quark in a simple model based on the NJL-jet model~\cite{Matevosyan:2010hh,Matevosyan:2011ey,PhysRevD.86.059904, Matevosyan:2011vj,Matevosyan:2012ga,Matevosyan:2012ms,Casey:2012ux,Casey:2012hg,  Matevosyan:2013nla, Matevosyan:2013aka}. In this model we use the NJL-jet framework to describe the quark hadronisation process, and allow for an elementary Collins effect in one of the hadron emission steps. We use MC simulations to extract both polarised single- and di-hadron FFs using their probabilistic interpretation. We study the possible sine modulations of these FFs with respect to the Collins, \vfr\, and $\vf_T$~angles (defined in the next section), respectively, in order to establish whether the elementary Collins effect can generate terms in polarised DiFFs with modulations that are expected to be induced by IFFs $H_1^\sphericalangle$ and $H_1^\perp$, as expected within the standard TMD approach~\cite{Bianconi:1999cd}. This information will be helpful in further developments of the state-of-the-art non-perturbative models for DiFFs~\cite{Bacchetta:2006un}. 
 
 This paper is organised in the following way.  In the next Section of this article we will briefly describe the details of the model used to extract the polarised FFs. In Section~\ref{SEC_RESULTS}  we will present the results for the single and dihadron FFs and we will finish with the conclusions in Section~\ref{SEC_CONCLUSIONS}.

\section{Simple Model for Hadronisation of a Transversely Polarised Quark}
\label{SEC_MODEL}  
   
\begin{figure}[!t]
\centering\includegraphics[width=1\columnwidth]{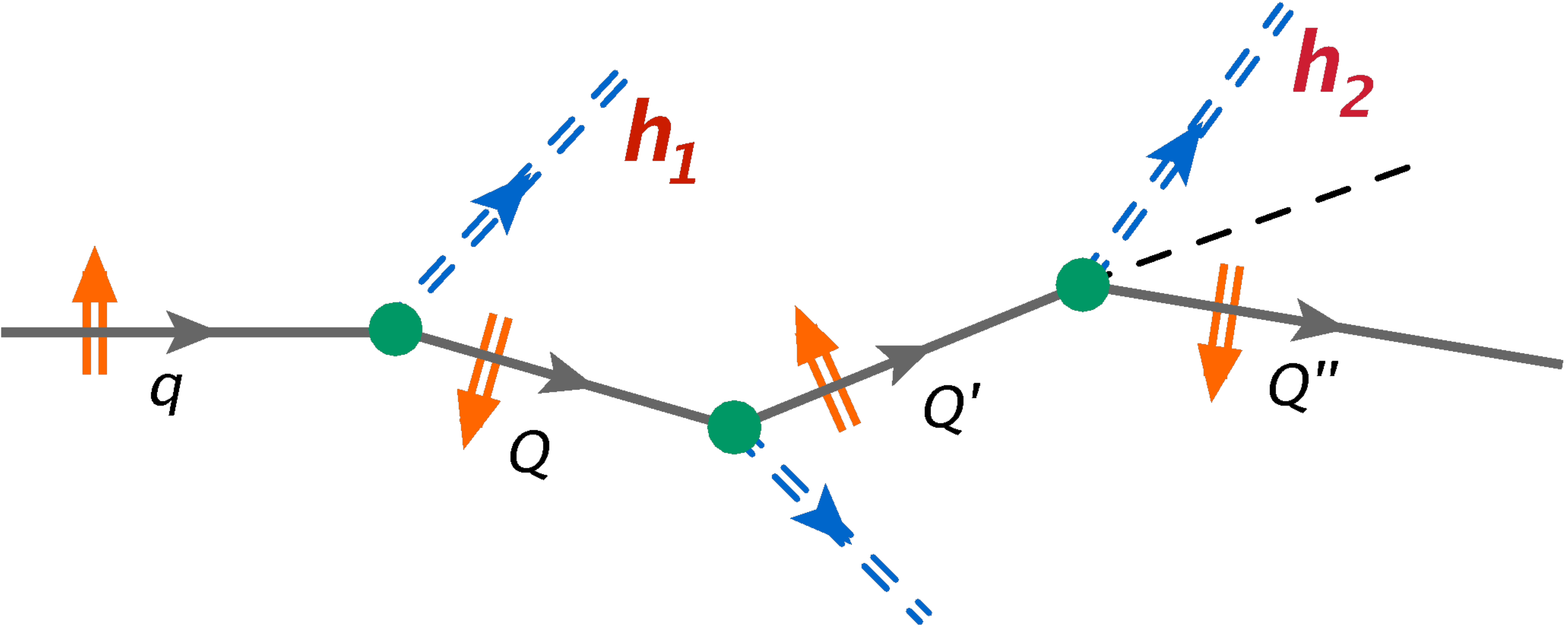}
\caption{NJL-jet model including transverse momentum and quark polarisation transfer. Here the orange double-lined arrows schematically indicate the spin direction of the quark in the decay chain.}
\label{PLOT_NJL-JET_TMD}
\end{figure}
 The NJL-jet model describes the quark hadronisation process within a framework based on the original Field and Feynman quark-jet picture~\cite{Field:1976ve,Field:1977fa}, where the initial fragmenting quark produces hadrons in a quark decay chain cascade, as schematically depicted in Fig.~\ref{PLOT_NJL-JET_TMD}. The remnant quark's properties after each hadron emission are determined using the flavour and momentum conservation constraints. In the NJL-jet model the elementary hadron emission probabilities at each vertex are calculated using the NJL  model.
 
 We first describe the kinematics and the MC method for calculating the single hadron FFs. This is followed by the calculation of dihadron FFs. 
 
 \subsection{Single Hadron Fragmentation Functions from  MC}

\begin{figure}[!tb]
\centering\includegraphics[width=1\columnwidth]{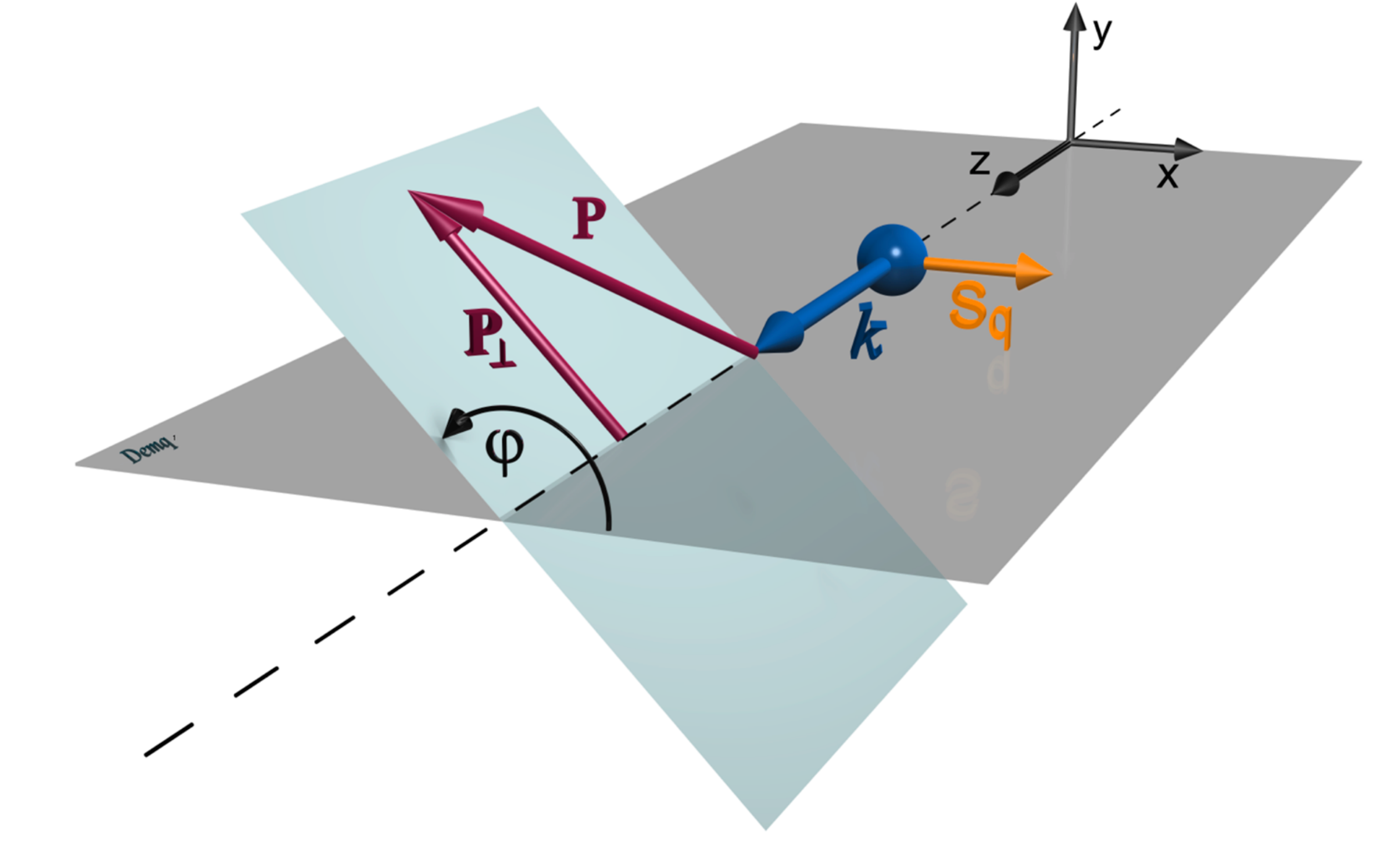}
\caption{Illustration of the three dimensional kinematics of transversely polarised quark fragmentation. 
The fragmenting quark's three-momentum $\vect{k}$ defines the $z$-axis with its transverse polarisation spin vector $\vect{S}_q$ along $x$ axis. The emitted hadron has momentum $P$ with the transverse component $\vect{P^{\perp}}$ with respect to the $z$-axis and azimuthal angle $\varphi$.}
\label{PLOT_POL_QUARK_3D}
\end{figure}

  We describe the single hadron FFs  in the quark-jet formalism as probability distributions for a quark to produce a hadron with certain properties. The relevant kinematics is schematically depicted in Fig.~\ref{PLOT_POL_QUARK_3D}.  The transversely polarised quark $q$  carries four-momentum ${k}$ and spin $\vect{S_q}$ and fragments to an unpolarised hadron $h$ of mass $m_h$ and four-momentum ${P}$.  The coordinate system is chosen such that the $z$ axis is along the direction of the three-momentum of the initial fragmenting quark $q$ and the $x$ axis is along the spin vector $\vect{S_q}$. Then the relevant momenta can be expressed as
\begin{align}      
\label{EQ_MOMENTA_FF}
&k=(k^-,k^+,\vect{0}),\
P = ( P^-,z k^+, \vect{P}^{\perp}),\ P^2 = m_{h}^2,
\end{align}
where $z\equiv P^+/k^+$ is the initial quark's light-cone momentum fraction carried by the hadron\footnote{We use the following LC convention for Lorentz 4-vectors $(a^-,a^+,\mathbi{a}^\perp)$, $a^\pm=\frac{1}{\sqrt{2}}(a^0\pm a^3)$ and $\mathbi{a}^\perp = (a^1,a^2)$. }. The Collins angle $\vf$ is taken as the angle between $\vect{P}^{\perp}$ and $\vect{S_q}$, as depicted in Fig.~\ref{PLOT_POL_QUARK_3D}.

  In the hadronization of a transversely polarised quark, the Collins effect describes a modulation of the unpolarised hadron fragmentation function with a  term proportional to the sine of the $\varphi$. This has been studied within the NJL-jet framework in Refs.~\cite{Matevosyan:2012ga, Matevosyan:2012ms}. Here we use the "Trento Convention" \cite{Bacchetta:2004jz} and the notation of Ref.~\cite{Matevosyan:2012ga} for the polarised fragmentation function
\begin{align}
\label{EQ_Dqh_SIN}
D_{h/q^{\uparrow}} (z,(P^\perp)^2,\varphi) &= D_1^{h/q}(z,(P^\perp)^2)\\ \nonumber
 &- H_1^{\perp h/q}(z, (P^\perp)^2) \frac{ P^\perp S_q}{z m_h} \sin(\varphi),
\end{align}
where the unpolarised fragmentation function is denoted $D_1^{h/q}(z,(P^\perp)^2)$ and $H_1^{\perp h/q}(z, (P^\perp)^2)$ is the  Collins function. 

The polarised FF can be integrated over $(P^\perp)^2$ and expressed in terms of the integrated unpolarised fragmentation function $D_1^{h/q}(z)$ and the $1/2$ moment of the Collins function $H_{1 (h/q)}^{\perp  (1/2)}(z)$
 \begin{align}
\label{EQ_Nqh_Phi}
D_{h/q^{\uparrow}}(z,\varphi) &\equiv \int_0^{\infty} d (P^\perp)^2\ D_{h/q^{\uparrow}}(z,(P^\perp)^2,\varphi) \\ \nonumber
=  \frac{1}{2\pi}&\left[D_1^{h/q}(z)\ - 2H_{1 (h/q)}^{\perp  (1/2)}(z) S_q \sin(\varphi) \right],
\end{align}
where
\begin{align}
\label{EQ_D1}
D_{1}^{h/q}(z) &\equiv \pi \int_0^{\infty} d (P^\perp)^2\ D_1^{h/q}(z, (P^\perp)^2),\\
\label{EQ_H12}
H_{1 (h/q)}^{\perp  (1/2)}(z) &\equiv \pi \int_0^{\infty} d (P^\perp)^2 \frac{P^\perp}{2z m_h}  H_1^{\perp h/q}(z,(P^\perp)^2).
\end{align}
 
  In the NJL-jet framework, the polarised fragmentation function is calculated as the MC average over $N_{Sims}$ simulations of the corresponding differential multiplicity, calculated with a fixed number of hadron emissions $N_L$ in each decay chain simulation
\begin{align}
\label{EQ_MC_EXTRACT}
&D_{h/q^{\uparrow}}(z, (P^\perp)^2,\varphi)\ \Delta z\ \frac{\Delta (P^\perp)^2}{2}\ \Delta \varphi \\ \nonumber
&= \left< N_{q^\uparrow}^h(z,z+\Delta z; (P^\perp)^2,(P^\perp)^2  +\Delta (P^\perp)^2; \varphi, \varphi+\Delta \varphi )\right>,
\end{align}
where $N_{q^\uparrow}^h$ is the number of hadrons of type $h$ produced by the quark $q$ that have momentum components within the regions specified in its arguments.  Our earlier studies of unpolarized FFs showed that the results of the simulations with different number of produced hadrons, $N_L$, rapidly converge  in any region of $z$ with an arbitrarily chosen lower edge (for example, the region with $z>0.2$ was saturated with just three hadron emissions in Fig.~5 of Ref.~\cite{Matevosyan:2011ey}).
  
  We take as input to these simulations the probabilities of emitting a single hadron, calculated with the NJL model~\cite{Matevosyan:2010hh,Matevosyan:2011ey,PhysRevD.86.059904, Matevosyan:2011vj}. In studies of the Collins function in Ref.~\cite{Matevosyan:2012ga}, the elementary Collins function used in every hadron emission step was also obtained using a quark model, where the relevant cut diagram involves a single gluon exchange with a gauge link. Here we will simplify the calculations by allowing only a single gluon is exchanged with an intermediate quark in the entire decay chain. Thus only a single hadron (chosen at random) in a given decay chain will be produced with the elementary Collins effect. Moreover, to highlight the mechanism for any modulations in the DiFFs induced by the Collins effect, we set the elementary Collins term (i.e. $H_1^{\perp h/q}(z, (P^\perp)^2) { P^\perp S_q}/{(z m_h)} $) in the polarised elementary fragmentation probability to be proportional to the unpolarised term ($D_1^{h/q}(z,(P^\perp)^2)$), with a constant coefficient of $0.9$. The idea is to highlight this particular effect and keep the polarised probability positive for all values of the corresponding arguments
\begin{align}
\label{EQ_MC_DRV_MOD}
d_{h/q^{\uparrow}} (z,(p^\perp)^2,\vf)=  d_1^{h/q}(z,(p^\perp)^2)(1-0.9 \sin({\varphi})),
\end{align}
where we assumed $S_q=1$ and $d_1^{h/q}$ is the elementary fragmentation function calculated from the NJL model. The above choice for the Collins term does not satisfy the constraint at (proportionality to)  vanishing transverse momentum dictated by the general arguments about angular momentum~\cite{Diehl:2005pc}. Nevertheless, this does not have a qualitative effect on the results presented here, as will be shown in our forthcoming work employing   the NJL-jet framework of Ref.~\cite{Matevosyan:2012ga} with Collins function calculated in the spectator model.

 The probability, $\SF$, of flipping the quark's spin orientation after each pseudoscalar hadron emission has been calculated in Ref.~\cite{Matevosyan:2012ga} using Lepage-Brodsky spinors in a helicity basis. The resulting $\SF$ depends on the magnitudes of the $x$ and $y$ components of the momentum of the emitted hadron, but is always bound $\SF>0.5$. Here, similar to our earlier model study in Ref.~\cite{Matevosyan:2012ms}, we will use constant values  of $\SF$ for simplicity. 

\subsection{Calculating DiFFs using MC}
\label{SUB_SEC_MC}

 Here we describe the kinematics of the fragmentation of a quark of flavour $q$ to two hadrons $h_1$ and $h_2$, following the standard conventions of Ref.~\cite{Radici:2001na}. The coordinate system is chosen such that the $z$ axis is along the direction of the 3-momentum of the initial fragmenting quark $q$.  The momenta of the quark and the two produced hadrons $h_1$ and $h_2$ are denoted as
\begin{align}      
\label{EQ_MOMENTA}
&k=(k^-,k^+,\vect{0}),\\ \nonumber
&P_{h1}\equiv P_1 = (P_1^-, z_1 k^+, \vect{P}_{1}^{\perp}),\ P_1^2 = M_{h1}^2, \\ \nonumber
&P_{h2}\equiv P_2 = (P_2^-, z_2 k^+, \vect{P}_{2}^{\perp}),\ P_2^2=M_{h2}^2,
\end{align}
where $z_1\equiv P_{h1}^+/k^+, M_{h1}$ and $z_2\equiv P_{h2}^+/k^+, M_{h2}$ are the corresponding light-cone momentum fractions and the masses of the hadrons. 

 The polarised DiFFs (PDiFF) depend on six kinematic variables that can be chosen, for example, as $z_1, z_2, P_1^\perp, P_2^\perp, \vf_1,\vf_2$, where the last four variables are the magnitudes and the azimuthal angles of $\vect{P}_1^\perp$ and $\vect{P}_2^\perp$. Often $\vect{P}_1^\perp$ and $\vect{P}_2^\perp$ are replaced by their linear combinations:
\al
{
\vect{R} = \frac{\vect{P}_{h_1}^\perp - \vect{P}_{h_2}^\perp}{2},\  \vect{P}_T = \vect{P}_{h_1}^\perp + \vect{P}_{h_2}^\perp,
}
with  corresponding azimuthal angles $\fr$~and $\vf_T$ (thus PDiFFs become functions of $z_1,z_2, R, P_T,\fr,\vf_T $).  Further, using parity and rotational invariance, PDiFFs can be decomposed into spin independent and spin dependent parts. The part of PDiFF proportional to the transverse component of the quark's spin has terms containing $\sin(\fr)$ and $\sin(\vf_T)$, as detailed in Refs.~\cite{Bacchetta:2003vn, Bianconi:1999cd}. We note that in our MC approach we have access to the full set of the variables. Here we will present our results as dependencies on (combinations) of some variables, while integrating over the others, similar to how the experimental data is presented. We will also explore $\sin(\fr)$ and $\sin(\vf_T)$ modulations of the PDiFFs in our model. 
 
 Let us as an example consider the unpolarised dihadron fragmentation functions $D_q^{h_1h_2}(z,M_h^2)$. They are functions of the sum of the light-cone momentum fractions $z=z_1+z_2$ and the invariant mass squared $M_h^2=(P_1+P_2)^2$, of the produced hadron pair. We employ the number density interpretation for the $D_q^{h_1h_2}(z,M_h^2)$ to extract them by calculating the corresponding multiplicities using a MC average over simulations of the quark hadronization process, similar to the method employed for the single hadron FF extractions~\cite{Matevosyan:2011ey,PhysRevD.86.059904,Matevosyan:2011vj,Matevosyan:2012ga}:
\begin{align}
\label{EQ_MC_DFF_EXTRACT}
D_q^{h_1h_2}(z,M_h^2)&\ \Delta z\ \Delta M_h^2 \\ \nonumber
&= \left< N_{q}^{h_1 h_2}(z,z+\Delta z; M_h^2,M_h^2  +\Delta M_h^2 )\right>,
\end{align}
where $ \left< N_{q}^{h_1 h_2}(z,z+\Delta z; M_h^2,M_h^2  +\Delta M_h^2 )\right>$ is the average number of  hadron pairs, $h_1 h_2$, created with total momentum fraction in range $z$ to $z+\Delta z$ and invariant mass squared in range $M_h^2$ to $M_h^2  +\Delta M_h^2$. This average is calculated over $N_{Sims}$ Monte Carlo simulations of the hadronization process.  For each MC simulation we consider all the hadron pairs produced  by the quark $q$ in a given decay chain and calculate their $z$ and $ M_h^2$, filling-in the corresponding histograms. In MC simulations we choose $N_{Sims}$~large enough and a sufficient number of discretization points for $\Delta z$, $\Delta M_h^2$ to avoid significant numerical errors. 

 We  calculate the DiFF of a transversely polarised quark analogously to the case of the single hadron FF using the number density interpretation:
 \al
 {\label{EQ_PDFF_DENS}
 &D_{\qp}^{h_1 h_2}(z, M_h^2, \fr)\ \Delta z\ \Delta M_h^2\ \Delta \fr 
 \\ \non
 &= \left< N_{\qp}^{h_1 h_2}(z,z+\Delta z; M_h^2,M_h^2  +\Delta M_h^2; \fr, \fr + \Delta \fr )\right>.
 }
 
  In general, the polarised DiFF contains one spin independent, $D_1$, and two spin dependent terms~$H_1^\sphericalangle$ entering with~$\sin(\varphi_R)$ modulation and~$H_1^\perp$  with~$\sin(\varphi_T)$ modulation. All these DiFFs depend on transverse momentum variables $R$, $P_T$  and $\vect{R}\cdot\vect{P}_T$~\cite{Bianconi:1999cd,Bacchetta:2003vn}. When integrated over $\vect{P}_T$ (as in Eq.~(\ref{EQ_PDFF_DENS})), the polarised DiFF will contain a sum of the unpolarised DiFF and a term involving only~$\sin(\varphi_R)$ modulation again, denoted by $H_1^\sphericalangle$ in ~\cite{Bacchetta:2003vn,Adolph:2014fjw}, but now the DiFFs depend only on one transverse variable, $R$. Note that, in general, both unintegrated  $H_1^\sphericalangle$ and~$H_1^\perp$  give contributions to the integrated $H_1^\sphericalangle$~\cite{Bacchetta:2003vn}.  Similarly, if integrated over $\vect{R}$, the polarised DiFF will correspond to a sum of the unpolarised DiFF and a term involving integrated~$H_1^\perp$ with~$\sin(\varphi_T)$ modulation.

The alternative choice for $\vect{R}$ introduced by Artru (see, e.g., Ref.~\cite{Artru:2002pua}, where the transverse momenta are weighted with light-cone momentum fractions of the other hadron), produce slightly different results (only significant for $\pi^- \pi^-$ pairs). We will present the analysis using the latter choice for $\vect{R}$ elsewhere.

\section{Monte Carlo Simulations and Results}
\label{SEC_RESULTS}

\begin{figure}[tb]
\centering 
\subfigure[] {
\includegraphics[width=\ImL]{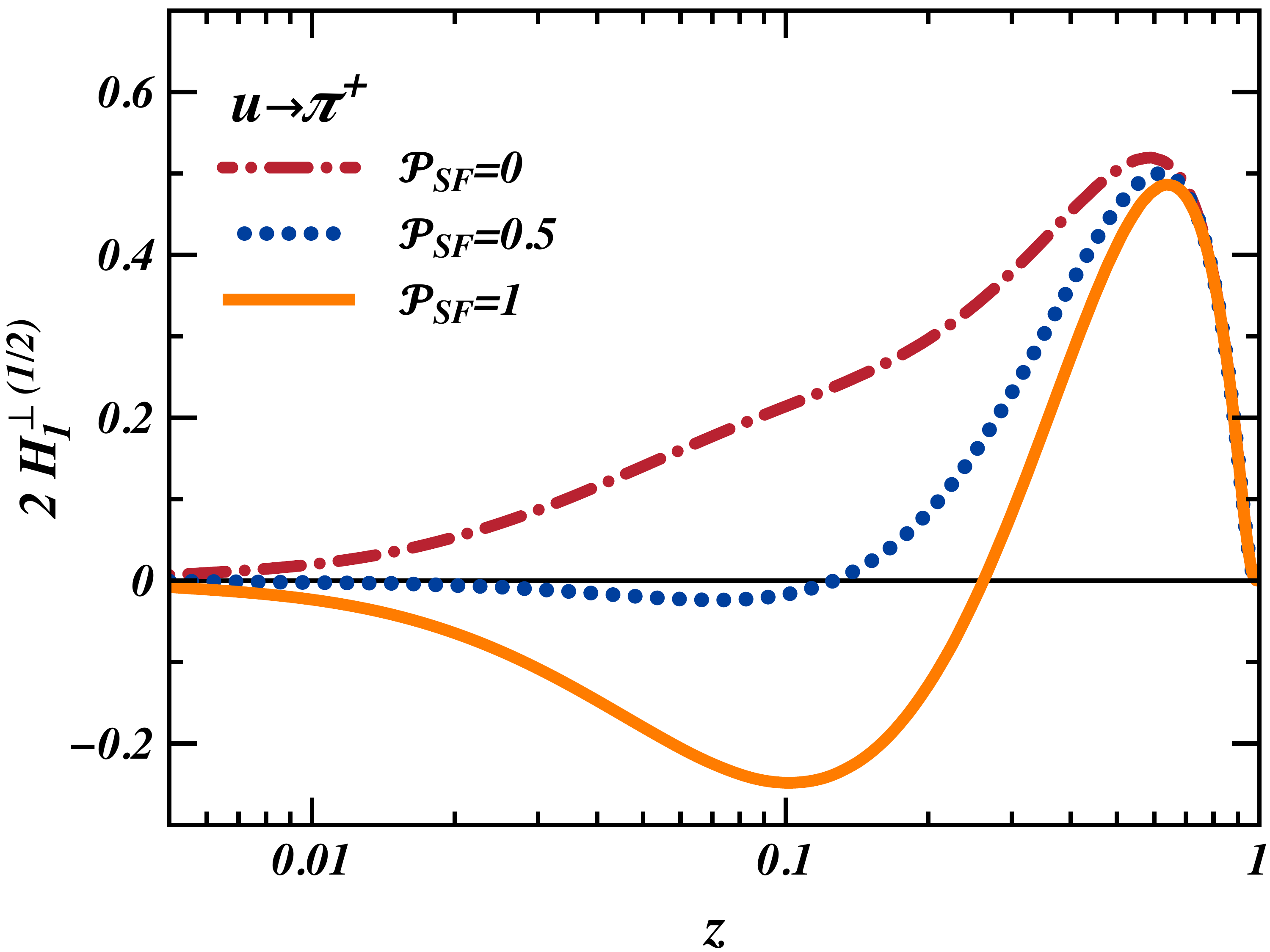}}
\\ 
\subfigure[] {
\includegraphics[width=\ImL]{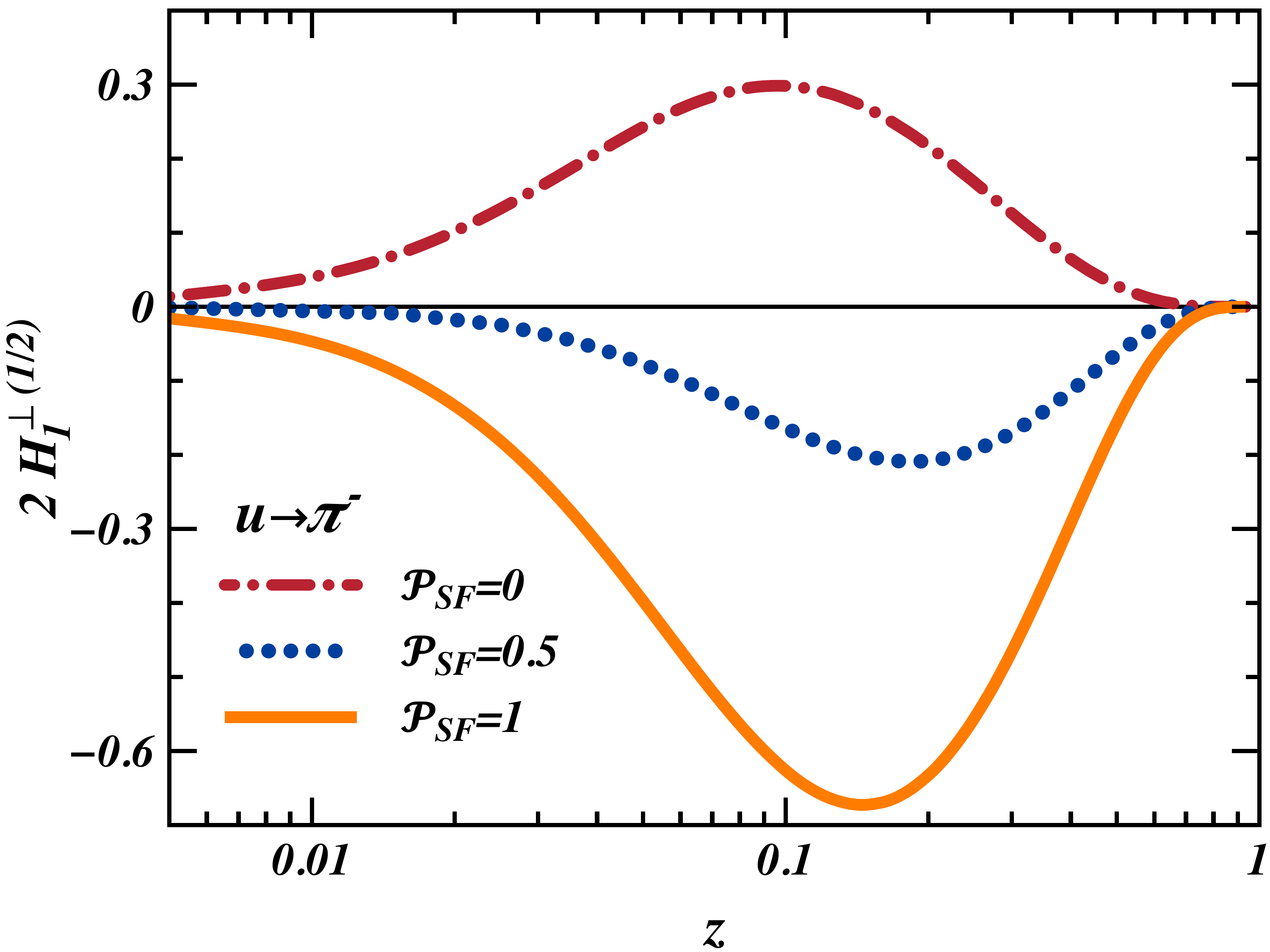}
}
\caption{Fitted values of $2 H_1^{\perp (1/2)}$ for $u\to\pi^+$ (a)  and $u\to\pi^-$ (b) as a function of $z$~from Monte Carlo simulations using three different values of $\SF$ at each elementary emission, where $N_L=2$.}
\label{PLOT_H12_PI}
\end{figure}
\begin{figure}[tb]
\centering 
\subfigure[] {
\includegraphics[width=\ImL]{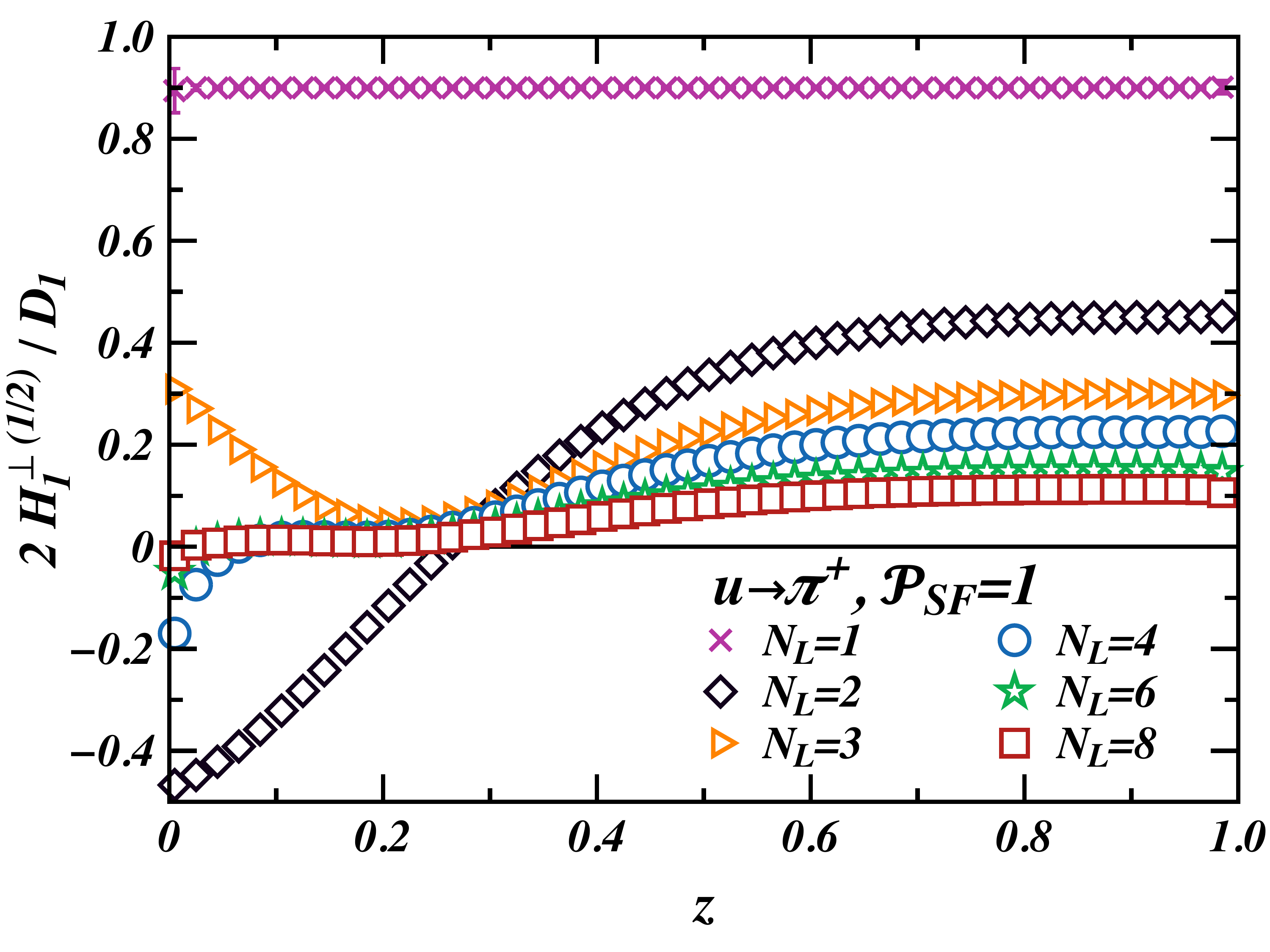}}
\\ 
\subfigure[] {
\includegraphics[width=\ImL]{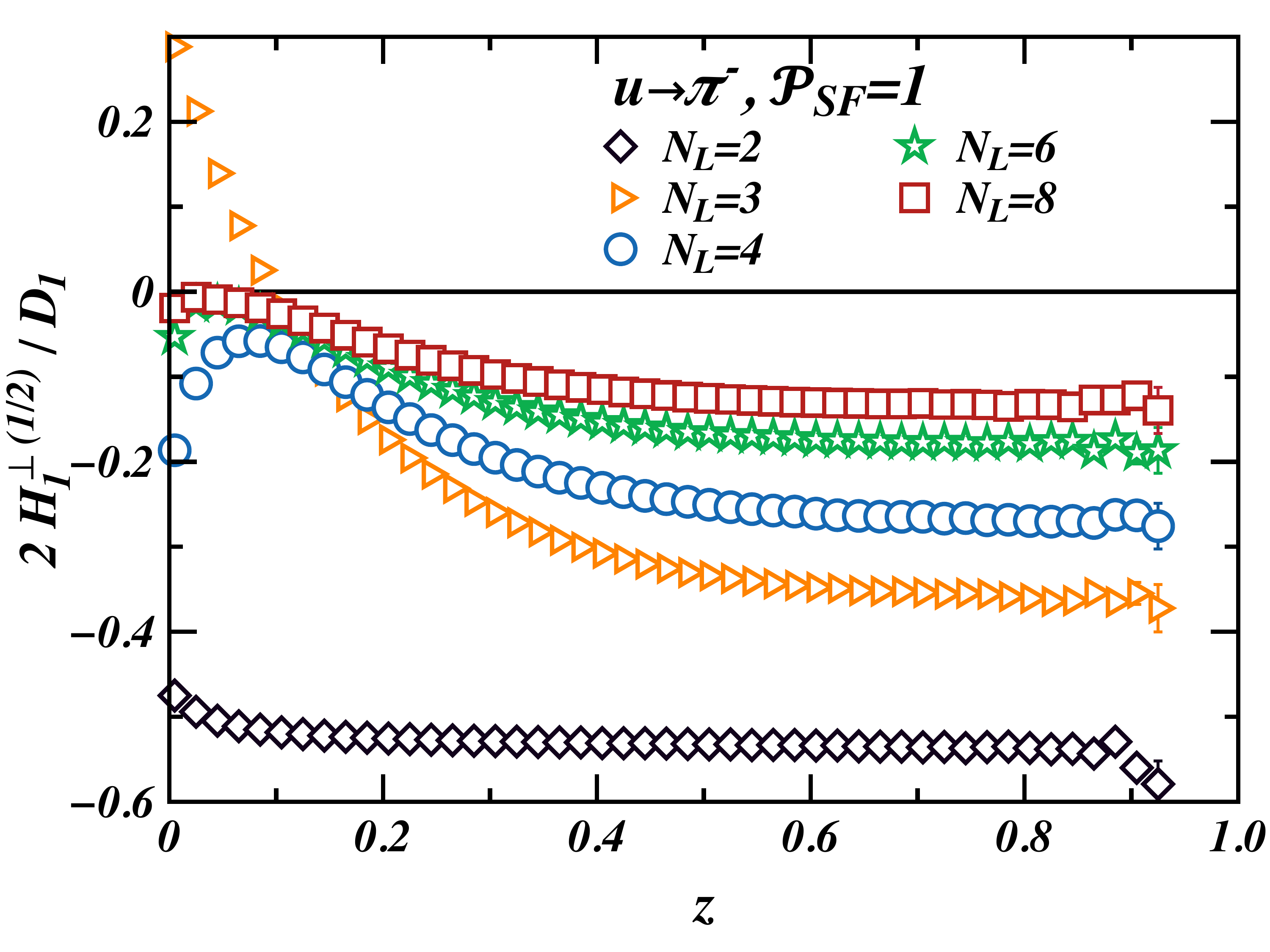}
}
\caption{Fitted values of the ratio $2 H_1^{\perp (1/2)}/D_1$ for $u\to\pi^+$ (a)  and $u\to\pi^-$ (b) as a function of $z$~from MC simulations with several values of $N_L$, where $\SF=1$.}
\label{PLOT_RatH12_PI}
\end{figure}

 Here we study the hadronisation of both transversely polarised and unpolarised $u$ quarks to pions, ignoring the production and decay of vector mesons, even though our recent studies of unpolarised DiFFs~\cite{Matevosyan:2013nla, Matevosyan:2013aka} showed their strong influence. We make this choice in order to focus on a particular piece of physics without extraneous complications. That is, we want to investigate whether the elementary Collins effect, together with the  quark spin flip mechanism, can produce spin dependent azimuthal modulations in two hadron production. We also ignore the strange quark and kaons in this calculation for the sake of the simplicity. As we have seen from our previous studies in the NJL-jet model~\cite{Matevosyan:2010hh,Casey:2012ux}, they simply change the overall normalisation of pion FFs. We performed MC simulations for a range of values for $N_L$ and three values of $\SF=\{0;0.5;1\}$ in the polarised case, choosing $N_{Sims}=10^{10}$. We discretized the invariant mass with $500$ points in the range up to $5~\Gs$ (starting from the two pion threshold), while the light-cone momenta fractions $z$ and all the angles are discretized with $100$ points. The parameters of the NJL model for the elementary splitting probability $ d_1^{h/q}(z,(p^\perp)^2)$ of Eq.~(\ref{EQ_MC_DRV_MOD}) are taken to be the same as in Ref.~\cite{Matevosyan:2011vj}.

\begin{figure}[tb]
\centering 
\subfigure[] {
\includegraphics[width=\ImS]{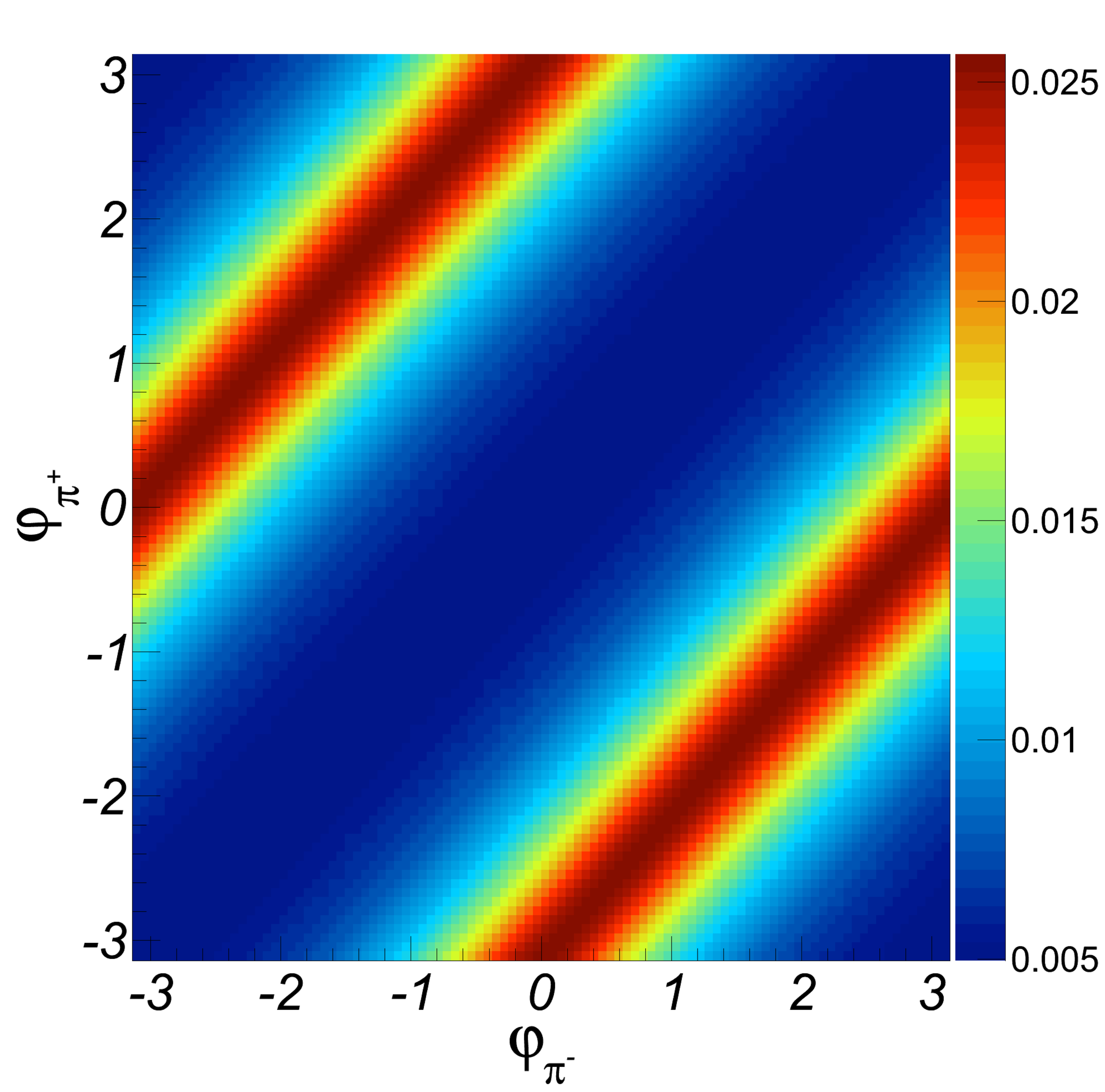}
}
\subfigure[] {
\includegraphics[width=\ImS]{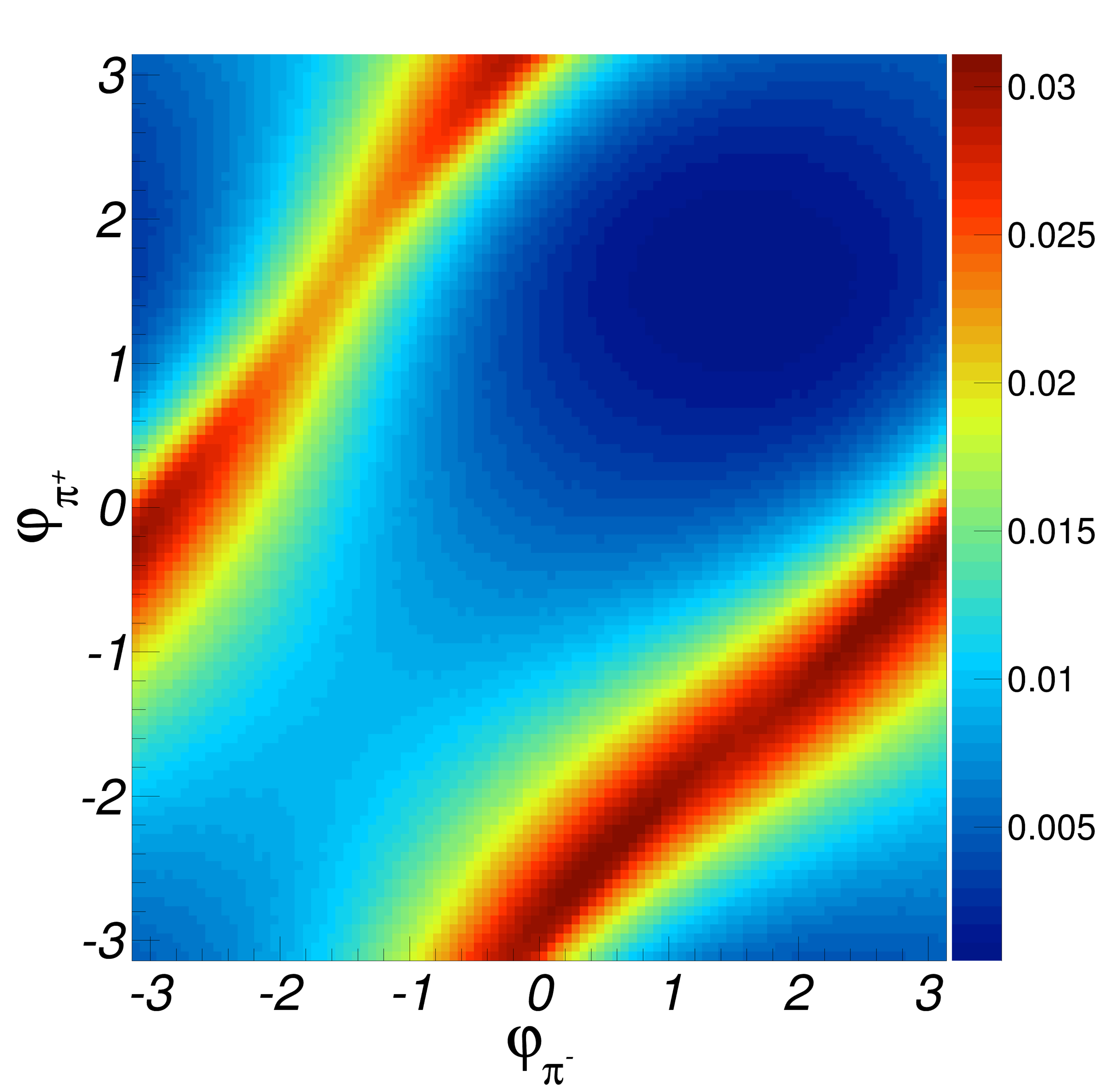}
}
\subfigure[] {
\includegraphics[width=\ImS]{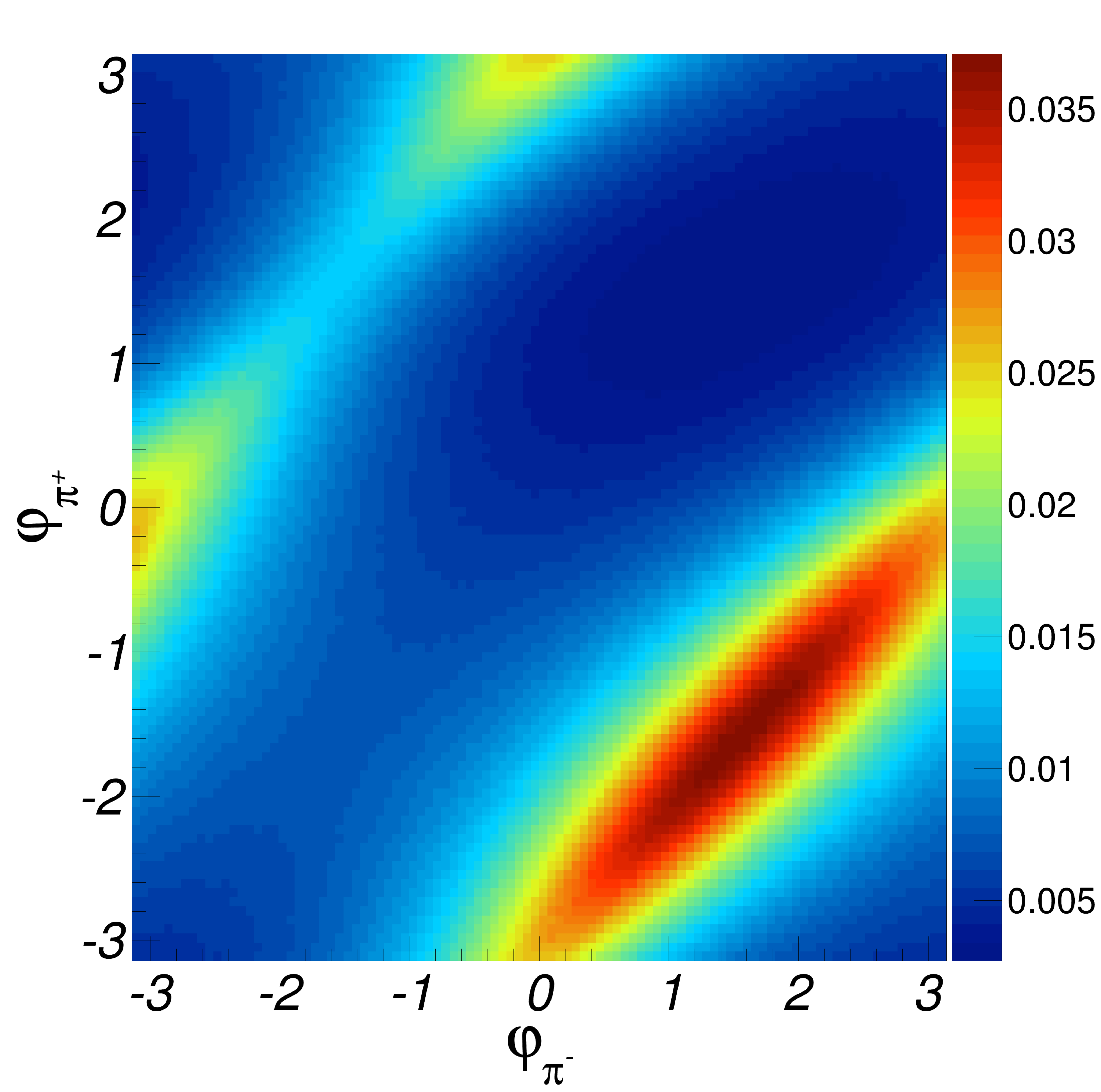}
}
\subfigure[] {
\includegraphics[width=\ImS]{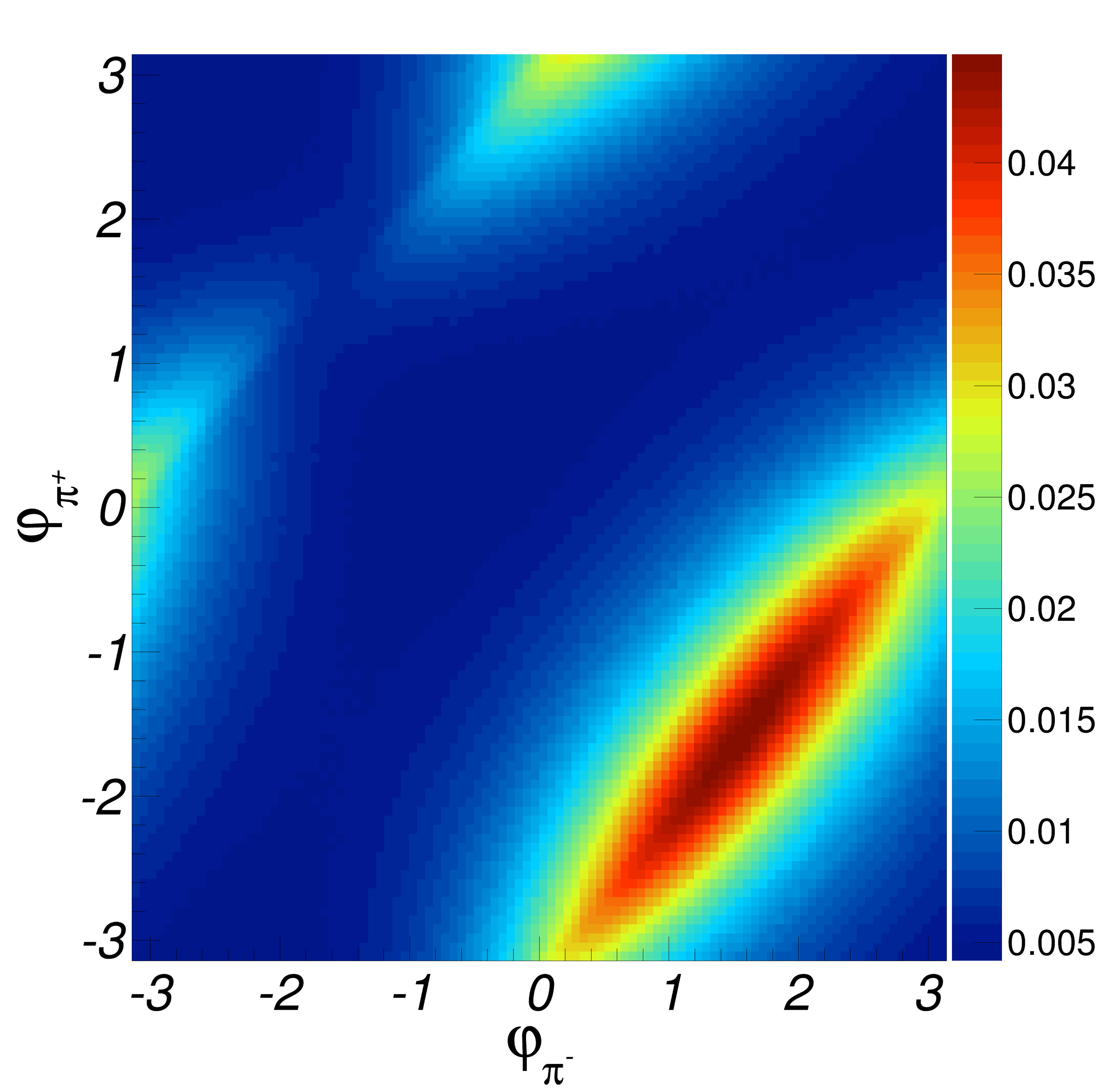}
}
\caption{The MC results for $D_{u^\uparrow}^{ \pi^+ \pi^-}(\varphi_{\pi^-}, \varphi_{\pi^+} )$ with unpolarised initial quark (a), transversely polarised quark with $\SF=0$ (b), $\SF=0.5$ (c) and $\SF=1$ (d), all with  $N_L=2$.}
\label{PLOT_PDFF_PHI1_PHI2}
\end{figure}
\begin{figure}[tb]
\centering 
\subfigure[] {
\includegraphics[width=\ImS]{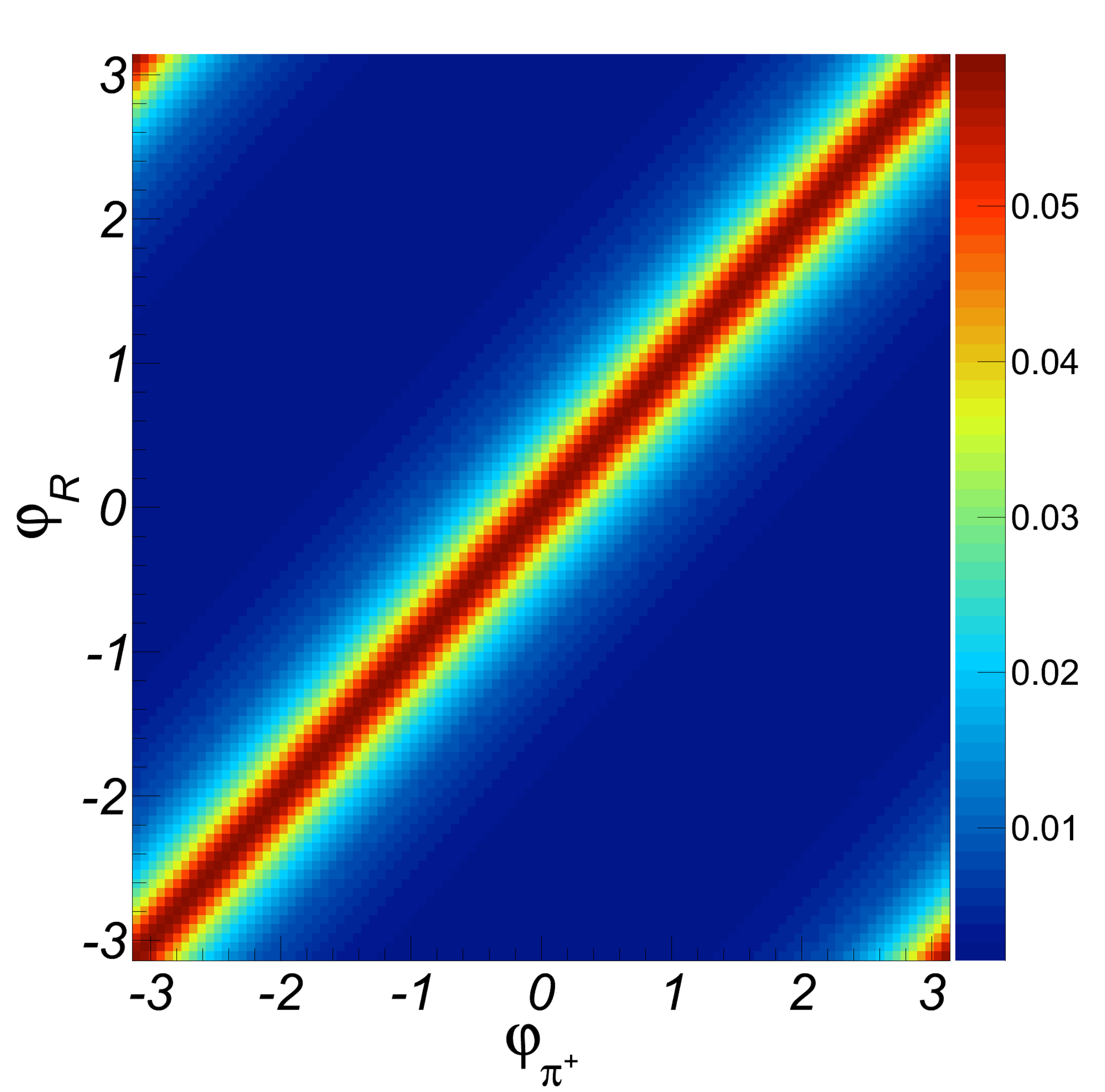}
}
\subfigure[] {
\includegraphics[width=\ImS]{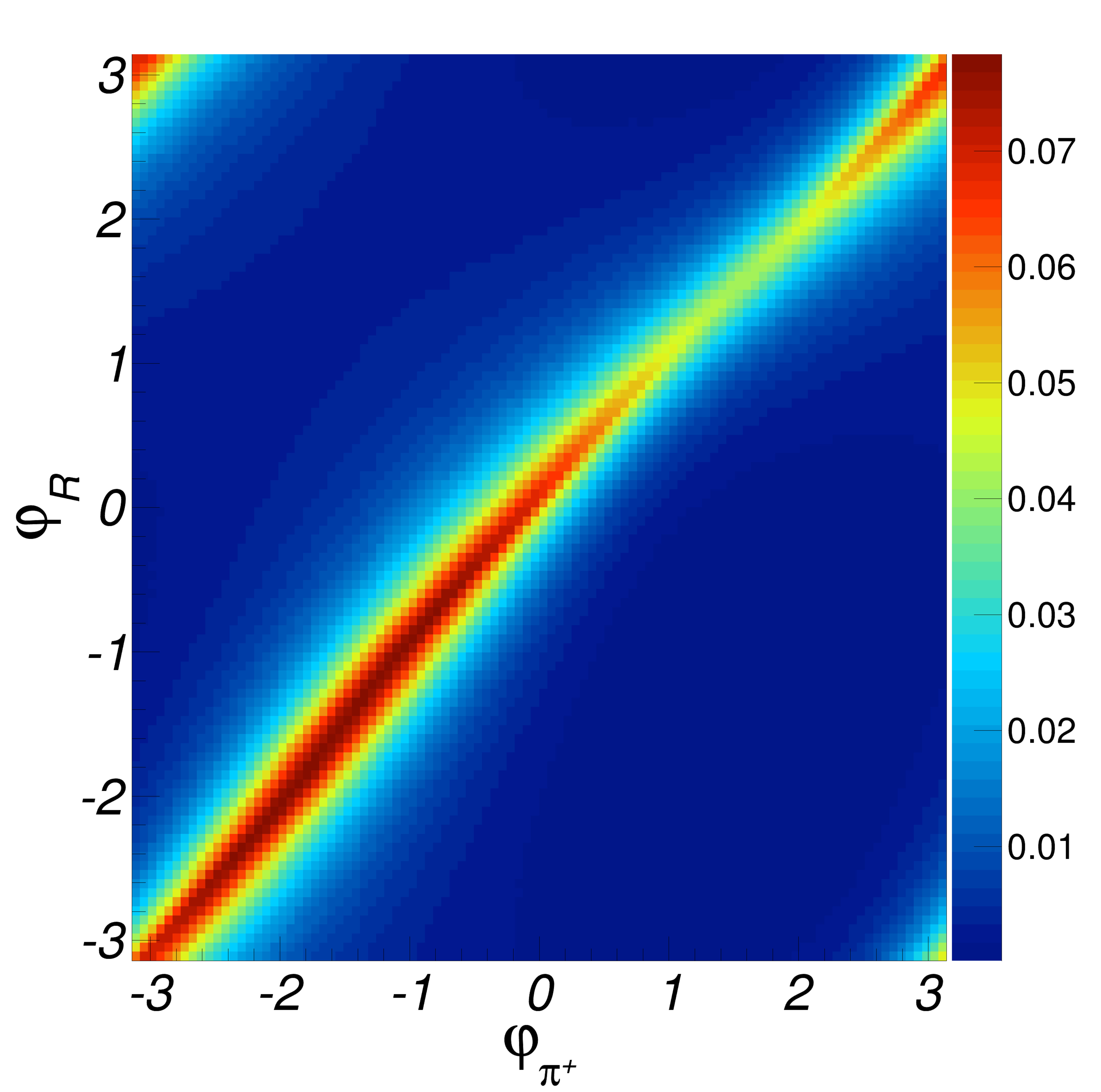}
}
\subfigure[] {
\includegraphics[width=\ImS]{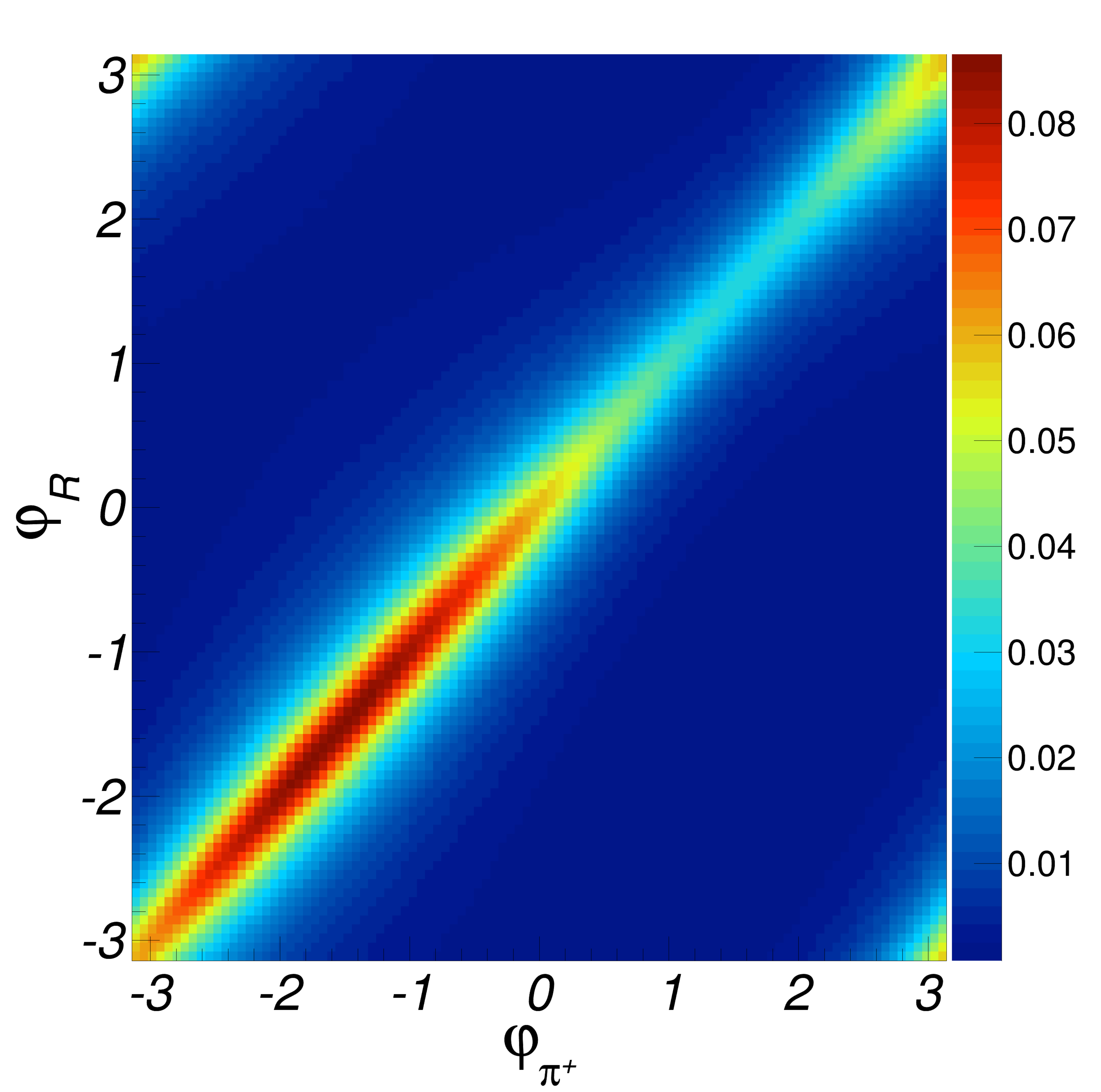}
}
\subfigure[] {
\includegraphics[width=\ImS]{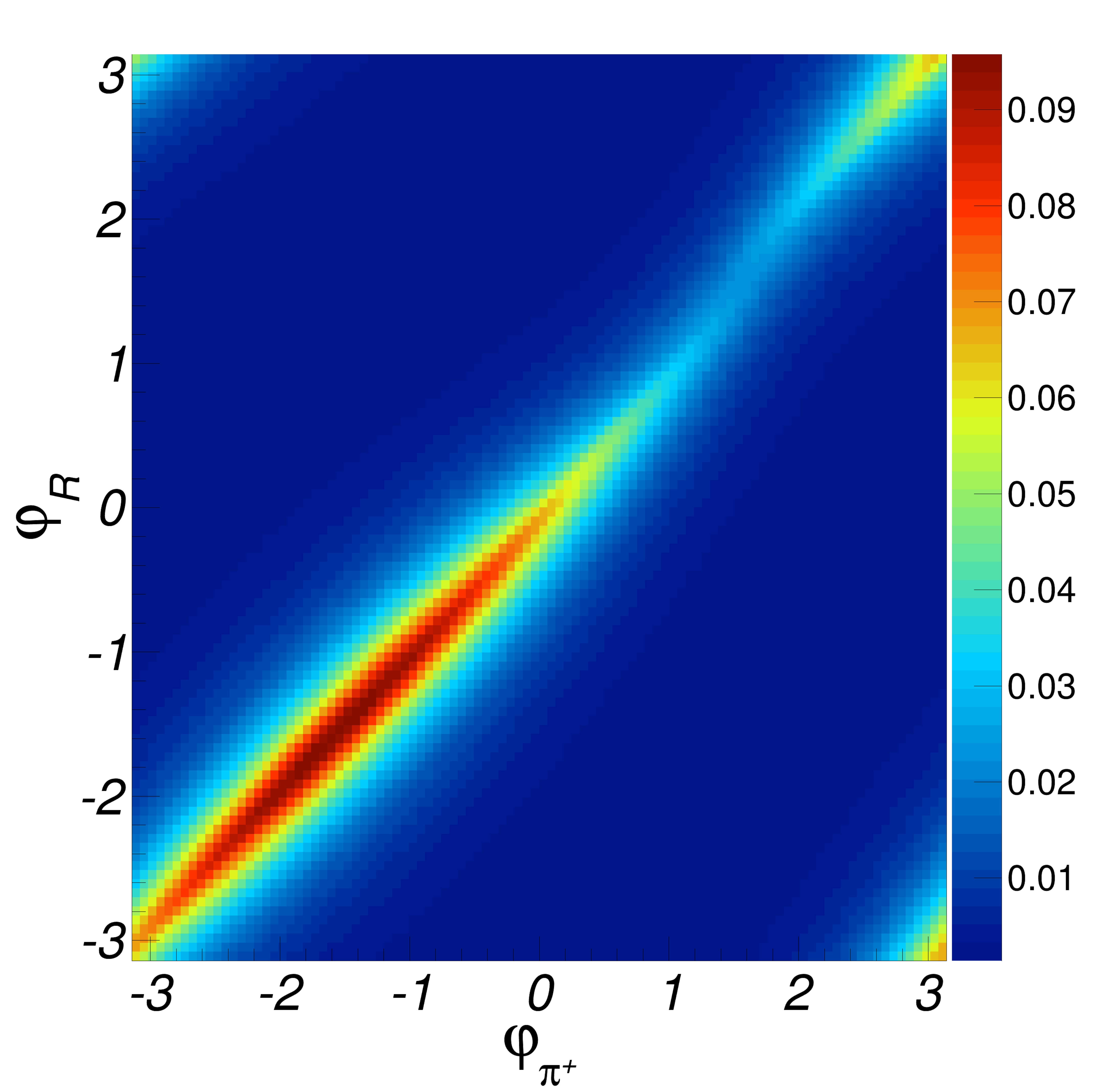}
}
\caption{The MC results for $D_{u^\uparrow}^{ \pi^+ \pi^-}(\varphi_{\pi^+}, \fr )$ with unpolarised initial quark (a), transversely polarised quark with $\SF=0$ (b), $\SF=0.5$ (c) and $\SF=1$ (d), all with $N_L=2$.}
\label{PLOT_PDFF_PHI2_PHIR}
\end{figure}

\subsection{Results for Collins Fragmentation Functions }
\label{SUBSEC_COLLINS}

  We first explore the dependence of the $1/2$ moments of the Collins functions on the quark spin flip probability $\SF$ by considering simulations with $N_L=2$, similar to our earlier work in Ref.~\cite{Matevosyan:2012ms} and using the same definitions. This allows us to isolate the quark spin flip mechanism's connection to the relative sign of the favoured and unfavored Collins functions. We extract the integrated unpolarised and the $1/2$ moment of the Collins fragmentation functions, using their relation to the $(P^\perp)^2$ integrated transversely polarised fragmentation.  We fit $D_{h/q^{\uparrow}}(z,\varphi)$ using a linear form
\al
{
   F(c_0,c_1) \equiv c_0-c_1 \sin(\varphi),
   \label{EQ_FIT_FUNC}
}
for fixed values of $z$, and identify the fitted coefficients $c_0,c_1$ with the unpolarised and Collins terms respectively. The performed fits work very well, with average $\chi^2$ per degree of freedom very close to unity. Once again, the extracted unpolarised fragmentation functions are unaffected by the quark spin flip probability $\SF$ and agree well with those studied previously in the NJL-jet model in Refs.~\cite{Matevosyan:2011ey,Matevosyan:2011vj}.
  
 The results from fits for the $1/2$ moments of the Collins function for fragmentation of a $u$ quark to $\pi^+$ and $\pi^-$, for three different value of $\SF$, are depicted in Fig.~\ref{PLOT_H12_PI}. Note that because of our choice of the elementary polarised fragmentation function in Eq.~(\ref{EQ_MC_DRV_MOD}), the shape of the $z$-dependence in this and following figures is governed by the quark-jet model dynamics and the $z$-dependence of elementary unpolarised fragmentation. These results agree qualitatively with those obtained in Ref.~\cite{Matevosyan:2012ms}. The first difference arises in the overall $1/2$ scaling of the Collins effect, which  that originates from the differences in models: here the elementary Collins effect occurs in only one hadron emission step, while in the model of Ref.~\cite{Matevosyan:2012ms} it occurs in every hadron emission step. The second difference concerns the choice of azimuthal angle of hadron in Eq.~(\ref{EQ_FIT_FUNC}): here the we use the azimuthal angle of the hadron transverse momentum with respect to the initial quark, while in Ref.~\cite{Matevosyan:2012ms} the azimuthal angle of the transverse momentum with respect to the quark directly producing the hadron was used. Here the fraction of the recoil transverse momentum of the quark transferred to the produced hadron at the second emission shifts the Collins functions.  The results for $\pi^+$ change with $\SF$ for $\NL=2$ simulations, as the contributions from those emitted at the first step are altered by those $\pi^+$ emitted in the second step (for the decay chains when the $u$ quark initially emits a $\pi^0$).  A $\pi^-$ on the other hand, can only be first emitted at $N_L=2$, after an initial $\pi^+$ emission, and is thus affected both by the spin flip and the recoil transverse momentum of the remnant quark. It is also remarkable, that unlike the results of Refs.~\cite{Matevosyan:2012ms,Matevosyan:2012ga}, the Collins effect for the $\pi^-$ does not vanish even when the quark depolarises ($\SF=0.5$), because of recoil effects. It is apparent, that opposite sign and similar magnitude Collins effects for the favoured and unfavored fragmentations (at relatively large $z$) are possible only for values $\SF>0.5$. Thus the opposite sign Collins function features of the full NJL-jet model results reported in Ref.~\cite{Matevosyan:2012ga},  where $\SF$ is a function of the kinematic variables and is always greater than $0.5$, qualitatively do not depend on the details of the particular input unpolarised and Collins elementary fragmentation functions, but rather represent the characteristics of the quark-jet model with preferential spin flip probability.
 
  The plots in Fig.~\ref{PLOT_RatH12_PI} depict the dependence of the fitted values for the analysing power of the Collins effect $2 H_1^{\perp (1/2)}/D_1$ (i.e. $c_1/c_0$ of the fit function in Eq.~(\ref{EQ_FIT_FUNC}) ) on $N_L$ for for both $u\to\pi^+$ (a)  and $u\to\pi^-$ (b). It is clear that the large $z$ behaviour of this ratio converges starting at $N_L=4$, modulo the $1/N_L$ scaling of the Collins effect in the current model (only one hadron emission step has an elementary Collins modulation in the entire decay chain.) Moreover, the ratio $2 H_1^{\perp (1/2)}/D_1$ is slightly larger for unfavored fragmentation, $u\to\pi^-$, than for the favoured process, $u\to\pi^+$.

\subsection{Results for Dihadron Fragmentations}
\label{SUBSEC_MC_RES_IFF} 

In this subsection we first present the results for $u\to \pi^+ \pi^-$, producing analogues of most of the plots presented by the COMPASS collaboration~\cite{Adolph:2014fjw,COMPASS:2013combo}.  Then we will study the azimuthal sine modulations of the DiFFs with respect to $\vf_R$ and $\vf_T$ for oppositely charged pion pairs. We further analyse the dependence of these modulations on the number of produced hadrons, $N_L$, in each quark decay chain. In this subsection $z$ denotes the sum of the light-cone momentum fractions of the hadron pair.

\subsubsection{Angular Modulations in $u\to \pi^+ \pi^-$ Fragmentation.}
\label{SUBSEC_MC_RES_NL2} 

 First we study various angular distributions for the $u\to \pi^+ \pi^-$ , similar to those recently presented by the COMPASS collaboration~\cite{Adolph:2014fjw,COMPASS:2013combo}. We use MC simulations with $N_L=2$ with both unpolarised and transverse polarised fragmenting quark. In the latter case, we also study the dependence on the quark spin flip probability $\SF$. The goal is to determine the role of the elementary Collins effect in various angular distributions. Considering only two hadron emissions in each decay chain, the initial up quark has only a single channel to produce two charged pions in the quark-jet model: $u\to d + \pi^+ \to u + \pi^- + \pi^+$. This allows us to readily interpret the results and avoid any possible entanglements with different production channels. Further, in this subsection we integrate PDiFFs over $z$ and $M_h^2$ in the entire region.
    
 We first consider the density plots of the DiFF versus the azimuthal angles of $\pi^+$ and $\pi^-$, depicted in Fig.~\ref{PLOT_PDFF_PHI1_PHI2}. We notice that there is a significant correlation between these angles in the unpolarised quark hadronisation, as seen in subfigure (a). The source of this correlation is that the remnant quark, after the emission of first $\pi^+$, acquires an opposite sign recoil transverse momentum. A fraction of this transverse momentum is passed to the second produced hadron $\pi^-$, thus the difference of the azimuthal angles $\vf_{\pi^+}$ and $\vf_{\pi^-}$ has a  peak at $\pi$ (the double distribution becomes completely uniform when the quark recoil effect is turned off). Such behaviour is also seen in the Lund string model incorporated in PYTHIA event generator, where the origin of this effect is named "local transverse momentum cancelation"~\cite{Artru:2002pua}. The elementary Collins effect distorts this picture because of the sine modulations of the hadron production probabilities with respect to their azimuthal angle, but critically does not change the distributions with respect to $\vf_{\pi^+}-\vf_{\pi^-}$.

The plots in Fig.~\ref{PLOT_PDFF_PHI2_PHIR} depict the PDiFF as a function of $\varphi_{\pi^+}$~and \vfr. The results for the unpolarised quark show that \vfr~follows the unintegrated azimuthal angle (here $\varphi_{\pi^-}$~is integrated implicitly through MC sampling). These simplistic correlations are distorted once the transverse polarisation of the quark is included. 

 The angular distributions presented here for an unpolarised quark resemble those presented by the COMPASS collaboration. The extreme parameters of the model (very large elementary Collins effect without kinematic zero at $p^\perp=0$ and endpoint values of $\SF$) exhibit a very strong variations from the unpolarised case. In a real-world scenario such variations might be too small to be easily seen in the experimental studies. 
\begin{figure}[tb]
\centering 
\includegraphics[width=\ImL]{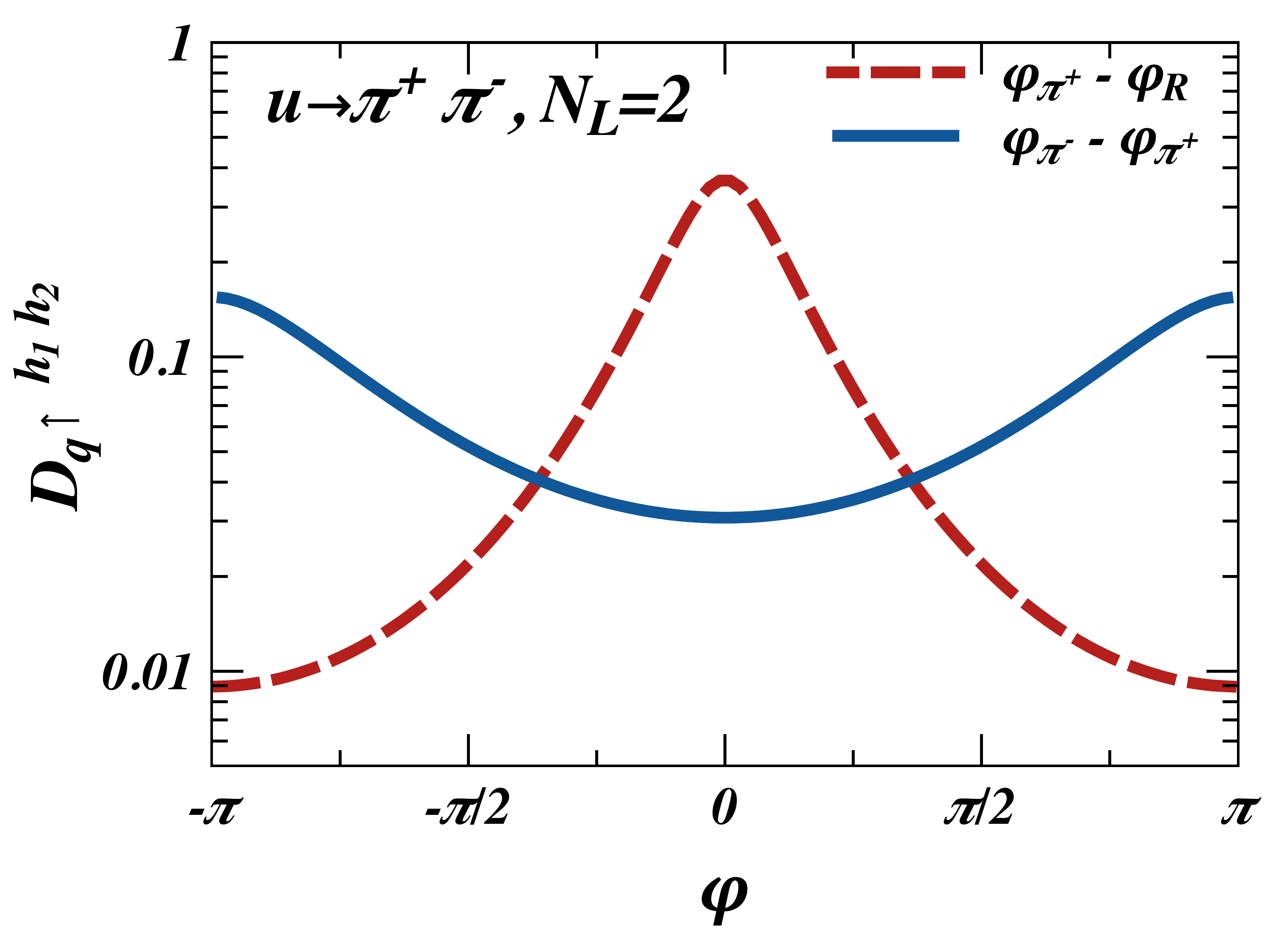}
\caption{The MC results for dependence of $D_{u^\uparrow}^{ \pi^+ \pi^-}$ on various angles, integrated over all the other variables. The initial quark is taken to be  unpolarised and $N_L=2$.}
\label{PLOT_PDFF_PHI_DIFF}
\end{figure}

 The plots in Fig.~\ref{PLOT_PDFF_PHI_DIFF} depict dependence of PDiFF on differences of various angles. A critical observation here is, that these results are not affected by the elementary Collins effect, but are purely generated by the kinematics of the dihadron production. 
 
\subsubsection{Dependence on \vfr~and $\vf_T$}

\begin{figure}[tb]
\centering 
\includegraphics[width=\ImL]{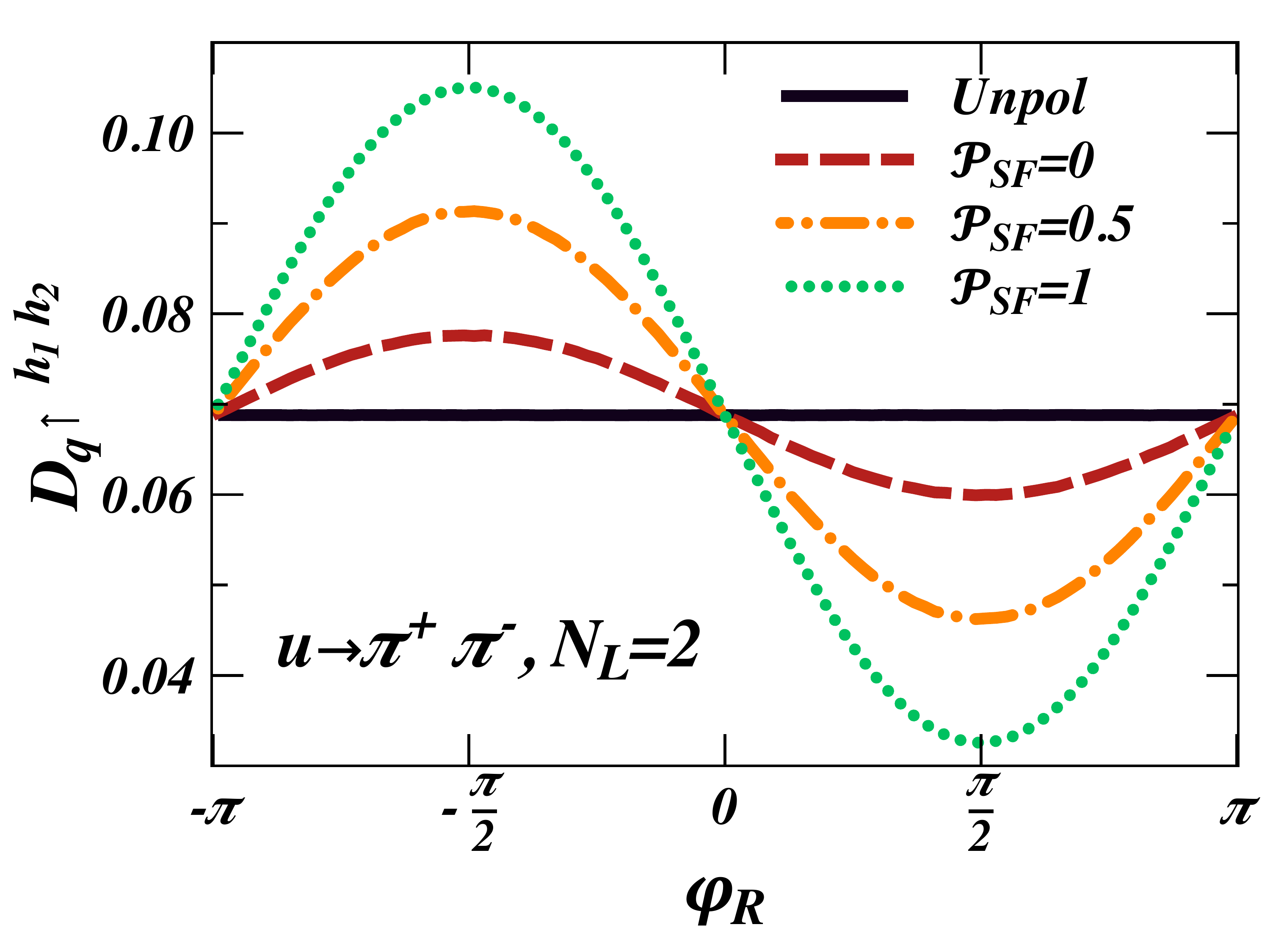}
\caption{The MC results for $D_{u^\uparrow}^{ \pi^+ \pi^-}(\fr )$ for $N_L=2$, integrated over all the other variables, with unpolarised initial quark (black solid line), transversely polarised quark with $\SF=0$ (red dashed line), $\SF=0.5$ (orange dash-dotted line) and $\SF=1$ (green dotted line).}
\label{PLOT_PDFF_PHIR_NL2}
\end{figure}
\begin{figure}[tb]
\centering 
\includegraphics[width=\ImL]{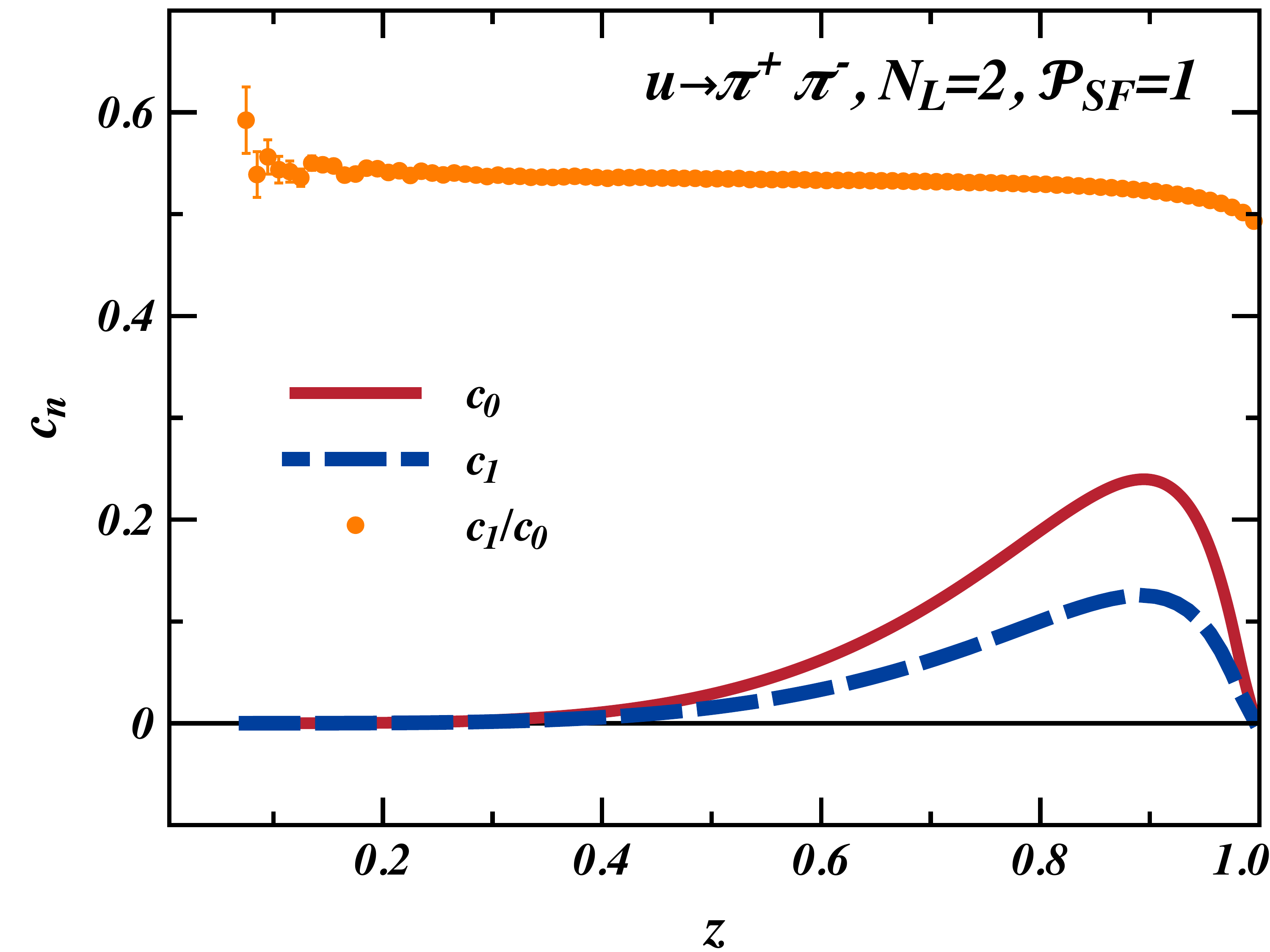}
\caption{The dependence of the parameters $c_0$, $c_1$ and $c_1/c_0$ versus $z$, fitted to the $\sin(\vf_R)$ modulation of PDiFF for $\SF=1$ and $N_L=2$.}
\label{PLOT_PDFF_PHIR_CN_NL2}
\end{figure}

 In this section we examine any possible sine modulations with respect to both of the angles \vfr~and $\vf_T$~in charged pion pairs. We first note that no modulations were found in our studies with an unpolarised initial quark. Thus any such signals for the transversely polarised quark are purely induced by the elementary Collins effect. 
 
 Figure~\ref{PLOT_PDFF_PHIR_NL2}, depicting $D_{u^\uparrow}^{ \pi^+ \pi^-}( \fr )$,  clearly shows~$\sin(\fr)$ modulations are present for $\pi^+\pi^-$ pairs for the polarised quark. The plots in Fig.~\ref{PLOT_PDFF_PHIR_CN_NL2} depict the results for the fitted parameters $c_0$ and $c_1$ of the function $F$ in Eq~(\ref{EQ_FIT_FUNC}) for a simulation with two emitted hadrons. Their ratio of $c_1/c_0\approx 0.5$  can be interpreted as half of the Collins effect in the elementary fragmentation of Eq.~(\ref{EQ_MC_DRV_MOD}), as only one of the produced pions has the elementary Collins modulation. 
 
\begin{figure}[tb]
\centering 
\includegraphics[width=\ImL]{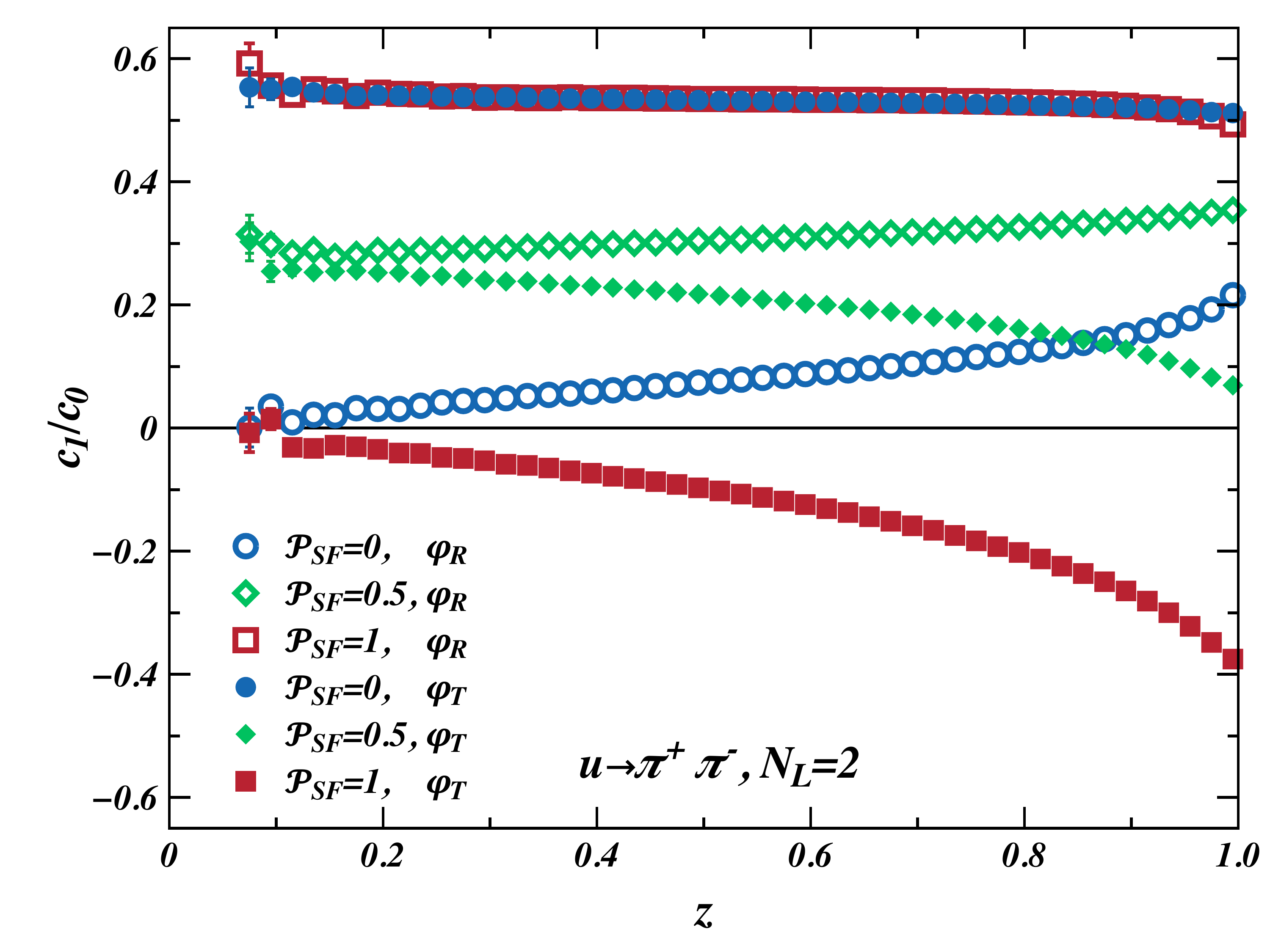}
\caption{The MC results for $c_1/c_0$ versus $z$ for $u\to\pi^+\pi^-$ fragmentation with $N_L=2$.  The asymmetries for $\sin(\fr)$~and $\sin(\varphi_T)$~are shown for three values of $\SF\in\{0,0.5,1\}$.}
\label{PLOT_PDFF_RAT_NL2}
\end{figure}

 The plots in Fig.~\ref{PLOT_PDFF_RAT_NL2} examine the dependence of both $\sin(\fr)$~and $\sin(\vf_T)$~modulations on $z$ for three values of $\SF$ in a simulations with two emitted hadrons. It is remarkable that both modulations are non-zero for all values of $\SF$. The features of these modulations for $\SF=0$ and $\SF=1$ can be understood by considering the preferential direction of the emitted hadrons in the transverse plane at every hadronization step. The traverse momentum $\vect{P}_{h_1}^\perp$ of the first emitted $\pi^+$ is preferentially pointing along  the $y$ axis (see Fig.~\ref{PLOT_POL_QUARK_3D}), since the elementary Collins function in Eq.~(\ref{EQ_MC_DRV_MOD}) is positive. The transverse momentum $\vect{P}_{h_2}^\perp$ of the $\pi^-$ emitted afterwards gets two contributions: a fraction of the recoil transverse momentum of the remnant quark after the first emission (pointing in the opposite direction to $\vect{P}_{h_1}^\perp$) and the relative transverse momentum with respect to the same quark. The second contribution preferentially points in the same or opposite directions to  $\vect{P}_{h_1}^\perp$  for $\SF=0$ and $\SF=1$, respectively (and becomes uniformly distributed for $\SF=0.5$). Thus the average of $\vect{P}_T$ for $\SF=0$ should point in the same direction as the average of $\vect{R}$ for $\SF=1$: along the $y$ axis. Thus the values of $c_1/c_0$ for the corresponding modulations are positive. The remaining results can also be easily interpreted using arguments along the same lines. 
 
\begin{figure}[!tb]
\centering 
\subfigure[] {
\includegraphics[width=\ImL]{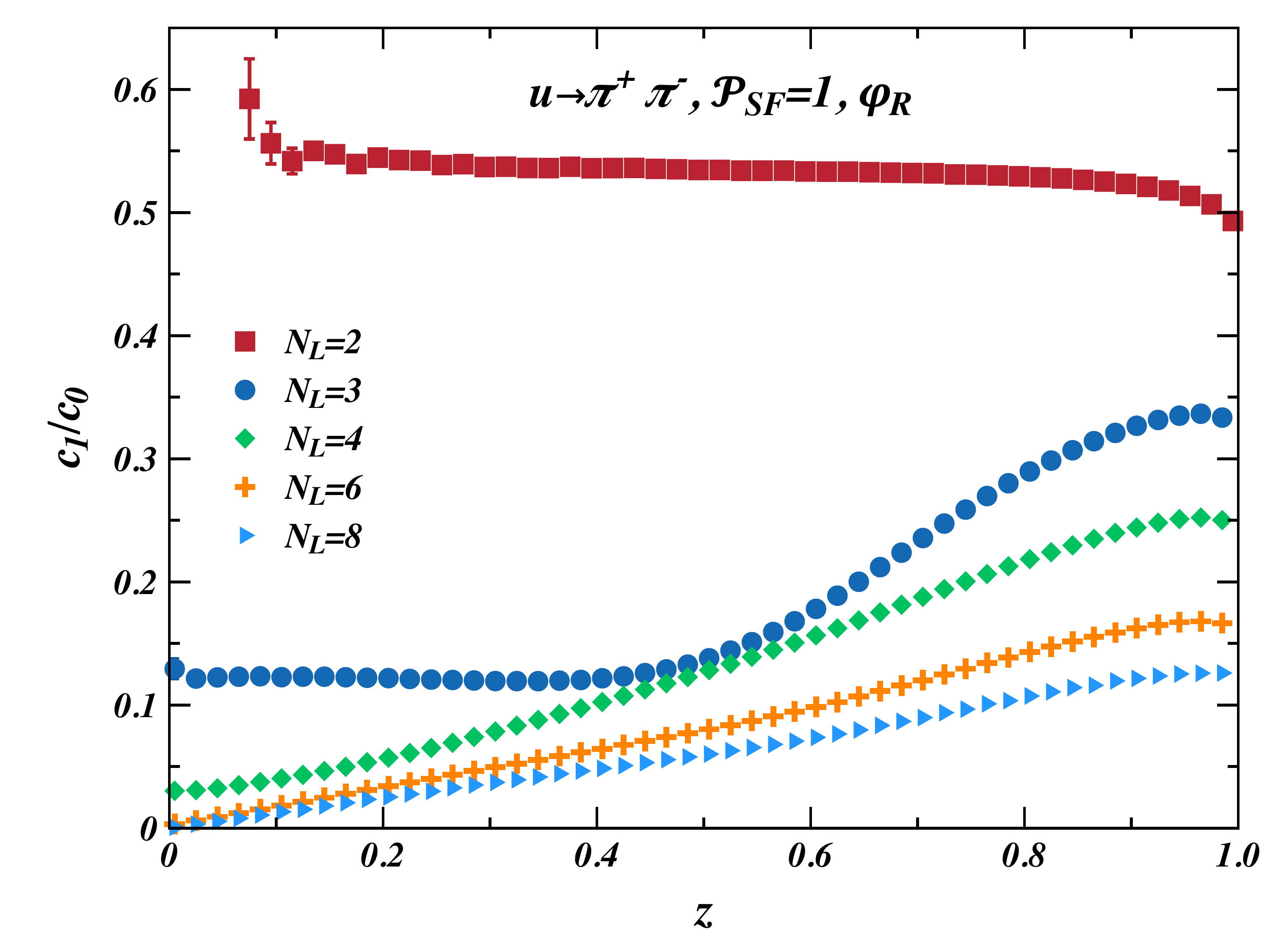}
}
\subfigure[] {
\includegraphics[width=\ImL]{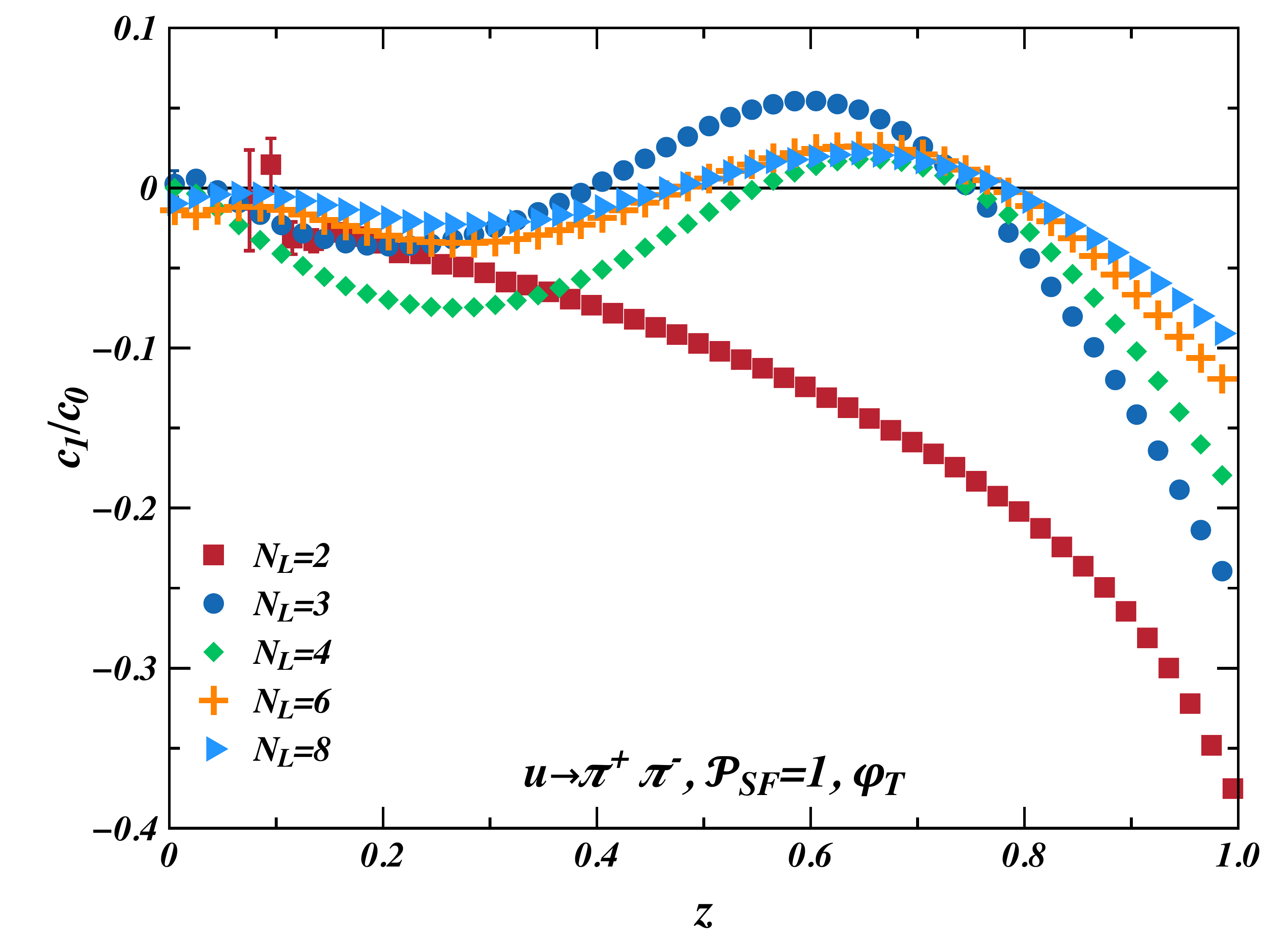}
}
\caption{The MC results for $c_1/c_0$ versus $z$ for $u\to\pi^+\pi^-$ fragmentation.  The dependence of the asymmetries for $\sin(\fr)$~and $\sin(\varphi_T)$~on $N_L$~are shown in subfigures (a) and (b) respectively.}
\label{PLOT_PDFF_RAT_NLX}
\end{figure}

The plots in Fig.~\ref{PLOT_PDFF_RAT_NLX} examine how these same sine modulations (\vfr~in subfigure (a) and $\vf_T$~in subfigure (b)) change with an increase of the number of emitted hadrons, when fixing $\SF=1$. In both cases, the large modulations at $N_L=2$ reduce and eventually converge (modulo $1/\NL$ scaling) for $\NL\gtrsim6$.  The large variations when $\NL$ changes from two to three can be explained by the fact that for $\NL=3$ two additional channels with intermediate $\pi^0$ emissions (that also flips the remnant quark's spin) can create $\pi^+\pi^-$ pairs with opposite signed modulation than the one involving  the first two emitted hadrons.

\begin{figure}[!tb]
\centering 
\subfigure[] {
\includegraphics[width=\ImM]{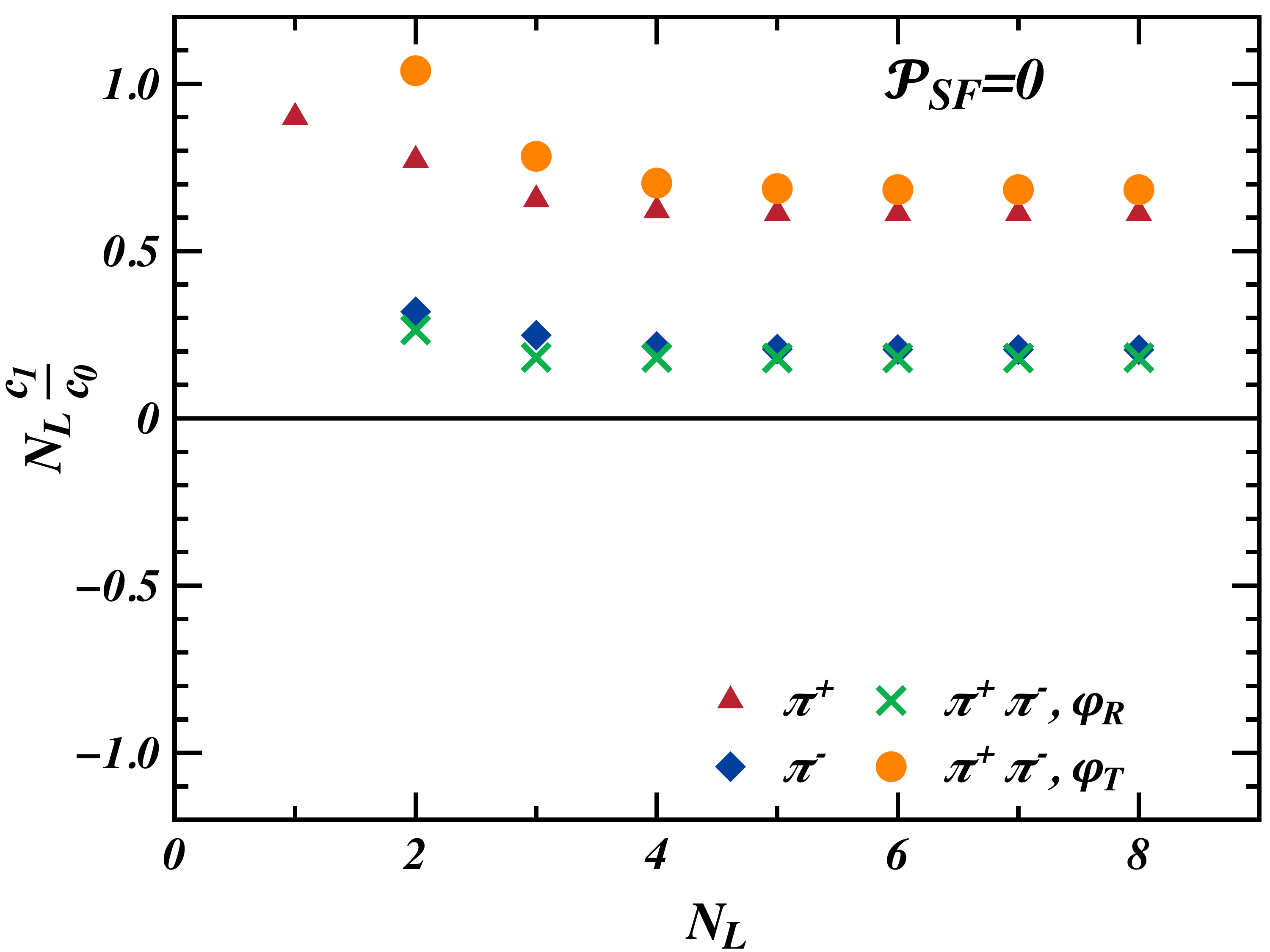}
}\\
\subfigure[] {
\includegraphics[width=\ImM]{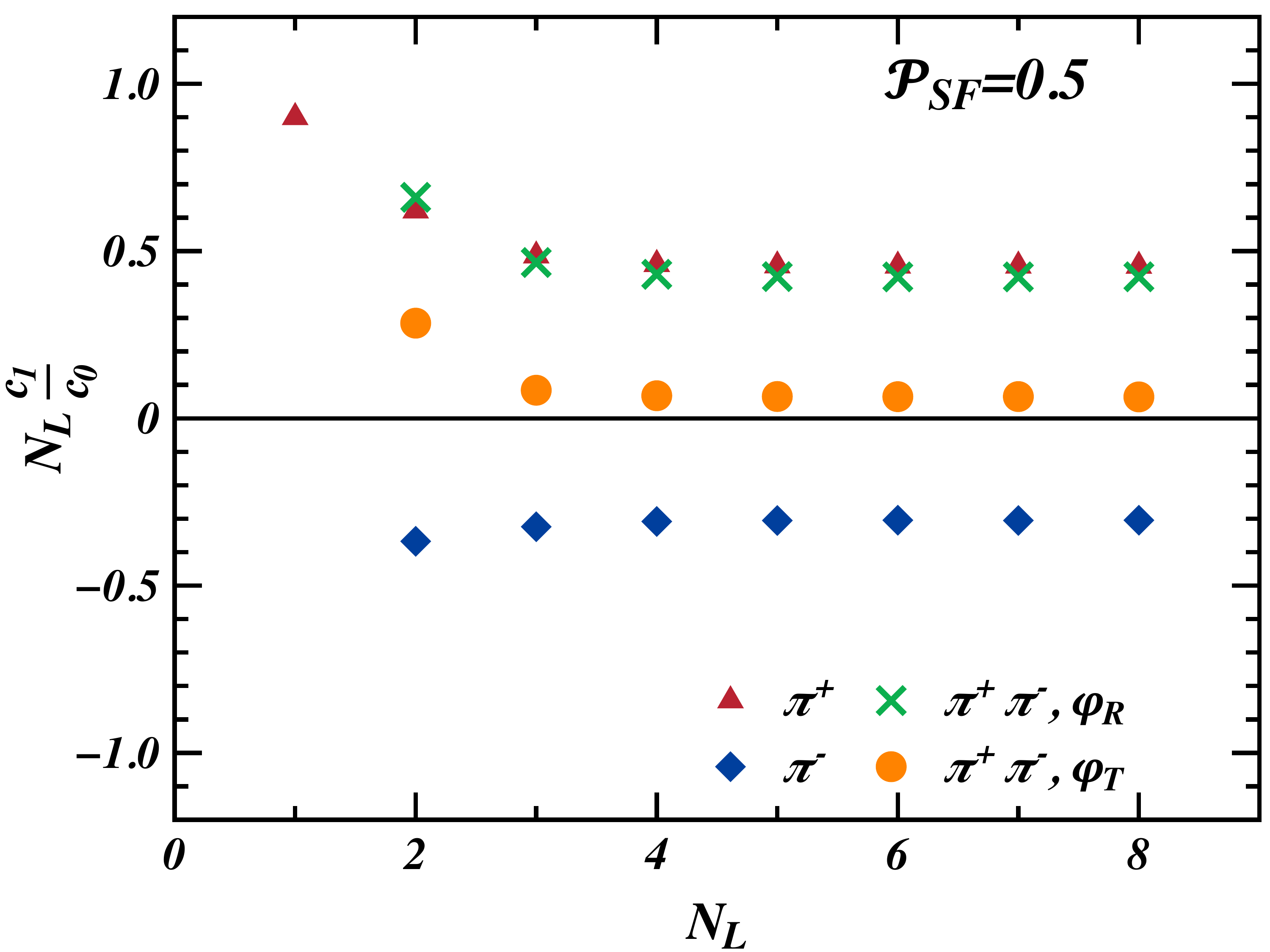}
}\\
\subfigure[] {
\includegraphics[width=\ImM]{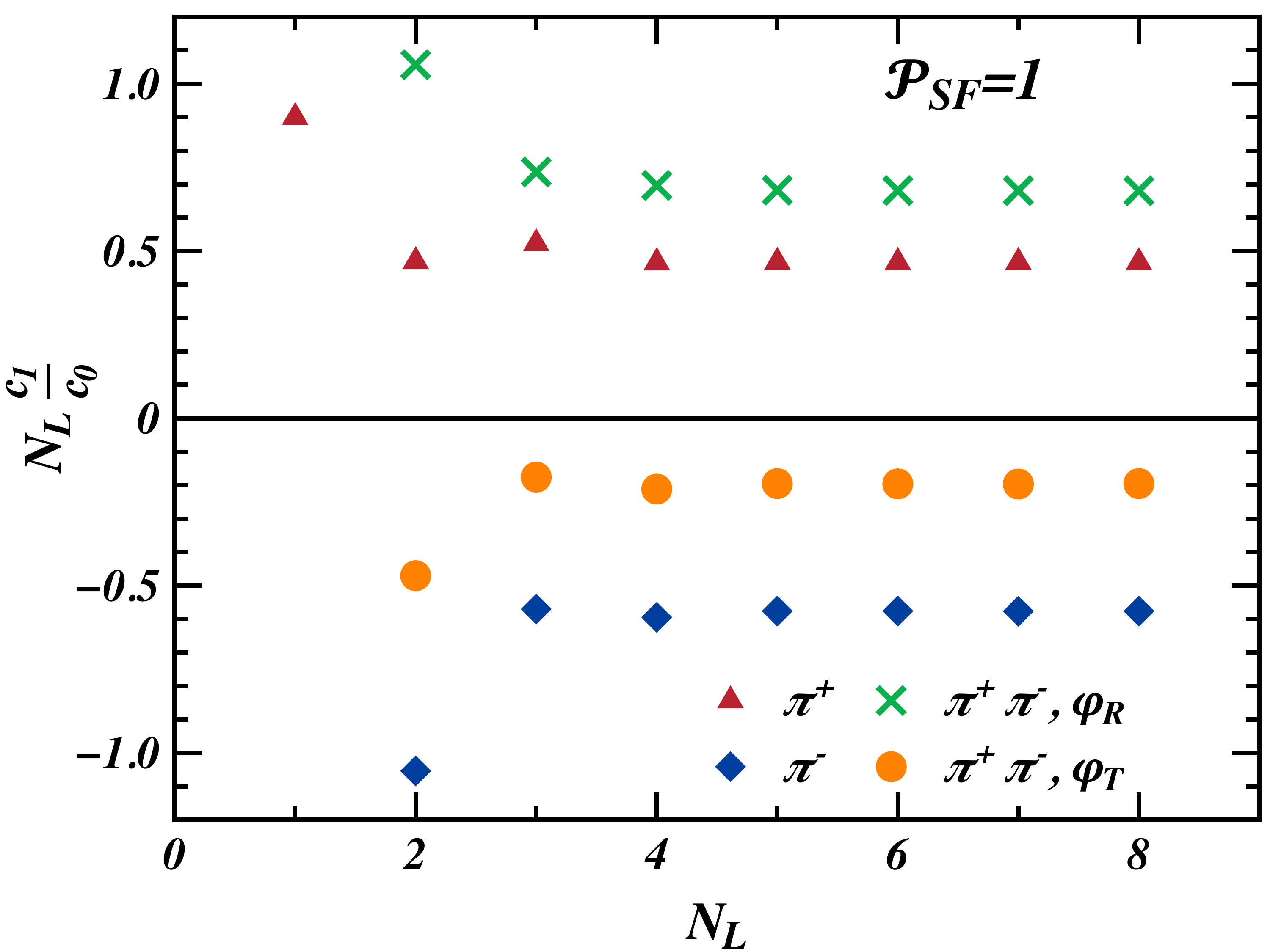}
}
\caption{
The MC results for the integrated, $N_L$-scaled analysing power $c_1/c_0$ versus $N_L$ for Collins effect for the charged pions and $\sin(\fr)$ and $\sin(\vf_T)$ modulations for their pairs versus $N_L$.  The results for three values of $\SF\in\{0,0.5,1\}$ are depicted in subfigures (a), (b) and (c) respectively. Here we impose a cut $z>0.1$ for each hadron. 
}
\label{PLOT_ANALYSE_PI_PI_NLX}
\end{figure}

Finally, Fig.~\ref{PLOT_ANALYSE_PI_PI_NLX} depicts the integrated analysing power (multiplied by $N_L$) of the Collins effect for the charged pions, and for  $\sin(\fr)$ and $\sin(\vf_T)$ modulations of $\pi^+\pi^-$ pairs versus $\NL$. Here the subfigures (a), (b) and (c) correspond to values of $\SF\in\{0,0.5,1\}$, respectively. We impose a cut on the light-cone momentum fraction of each hadron $z_{1,2}>0.1$ (thus $z>0.2$ for the pair), similar to that applied in the COMPASS analysis.  It is remarkable that for $\SF\geq 0.5$, the values of the analysing power of the $\sin(\fr)$~modulations are very close to those for the Collins effect for the $\pi^+$, while the analysing power of the Collins effect for the $\pi^-$ has the opposite sign and a similar magnitude.  These integrated analysing powers enter in products or convolutions in the expressions for the corresponding SSAs in SIDIS measurements.  Behaviour similar to those seen here for analysing powers have been seen in the results for the Bjorken $x$ dependence of the SSAs presented  by COMPASS~\cite{Adolph:2014fjw,COMPASS:2013combo}. Thus, assuming $u$ quark dominance in the large $x$ region, we can conclude that the similarities between the Collins and \vfr~SSAs can be explained by the simple quark-jet dynamics.

\section{Conclusions}
\label{SEC_CONCLUSIONS}

 In this work we explored the two hadron fragmentation functions interpreted as number densities for fragmentation of a quark with various assumptions for the spin dynamics in the quark-jet fragmentation. In particular, we considered the fragmentation of an unpolarised initial quark, as well as a polarised initial quark with different endpoint values for the quark spin flip after every hadron emission ($\SF=\{0;0.5;1\}$). We also allowed for a very large elementary Collins effect in each hadron emission step, while ensuring  that the positivity for the emission probability was not violated. We calculated  PDiFFs as a function of various arguments using the MC technique within the NJL-jet model formalism. Our results showed that the inclusion of the elementary Collins effect can produce Interference Fragmentation Function type $\sin(\fr)$ modulations of the  PDiFF for all values of the spin flip probability.  Moreover, for values of $\SF \geq 0.5$ the similarities between both the magnitudes and the signs of the analysing powers for Collins and $\sin(\fr)$ modulations appear to be akin to those for the corresponding SSAs recently reported by  COMPASS~\cite{Adolph:2014fjw,COMPASS:2013combo}. In our previous studies we showed that preferential spin flip probability ($\SF\geq0.5$) is also warranted if the Collins functions for an unfavored fragmentation has the opposite sign to that for the favoured one, which is strongly suggested by the experimental results of COMPASS~\cite{Bradamante:2011xu}, HERMES~\cite{Airapetian:2004tw}, JLab~\cite{Aghasyan:2011ha},  BELLE~\cite{Abe:2005zx,Seidl:2008xc}. Moreover our NJL-jet model calculations of the spin-flip probability of the quark also impose $\SF>0.5$. Finally, we also found that both $\sin(\fr)$ and $\sin(\vf_T)$ modulations are induced in all pairs of (both charged and neutral) pions, but we chose to only include here the results for $\pi^+\pi^-$ for brevity.

  In this work we only considered light quarks and direct pion emissions. The inclusion of the strange quark and kaons will only change the relative magnitudes of the PDiFFs, but will not change the effect itself, as we have observed in our previous studies. The inclusion of the production and decay of the vector mesons may be more important, as in the case of the unpolarised DiFFs. This shall be the subject of further investigation.
  
\section*{Acknowledgements}
 
This work was supported by the Australian Research Council through Grants No. FL0992247 (AWT), No. CE110001004 (CoEPP), and by the University of Adelaide.


\end{document}